\documentclass[10pt,twocolumn,superscriptaddress,prl,showpacs,floatfix,aps,preprintnumbers, nofootinbib]{revtex4-2}

\usepackage[utf8]{inputenc}

\usepackage{booktabs}
\usepackage{graphicx}
\usepackage{amssymb}
\usepackage{mathtools}
\usepackage{bm}
\usepackage{bbm}
\usepackage{xfrac}
\usepackage{multirow}
\usepackage{enumitem}
\usepackage{float}
\usepackage[dvipsnames]{xcolor}
\usepackage{etoolbox}
\usepackage{environ}
\usepackage{subcaption}
\usepackage{amsthm}
\usepackage{bibunits}
\usepackage{empheq}

\makeatletter
\newcommand{\setbibunitnamespace}[1]{%
  \gdef\@extra@binfo{#1}%
  \gdef\@extra@b@citeb{#1}%
}
\patchcmd{\NAT@bibsetnum}
  {\ref{LastBibItem}}
  {\ref{LastBibItem\@extra@b@citeb}}{}{}
\patchcmd{\endthebibliography}
  {\label{LastBibItem}}
  {\label{LastBibItem\@extra@b@citeb}}{}{}
\makeatother

\usepackage{ragged2e}
\usepackage[colorlinks=true]{hyperref}
\definecolor{linkblue}{rgb}{0.2,0.5,1}
\hypersetup{
    linkcolor = linkblue,
    citecolor = linkblue,
    urlcolor  = linkblue
}

\definecolor{draftcol}{HTML}{4A90D9}
\definecolor{commentblue}{rgb}{0.0,0.0,0.8}
\definecolor{commentteal}{rgb}{0.0,0.45,0.45}
\definecolor{commentorange}{rgb}{0.9,0.45,0.0}
\definecolor{commentmagenta}{rgb}{0.8,0.0,0.8}

\newtoggle{showdraft}
\toggletrue{showdraft}
\makeatletter
\renewcommand{\vec}[1]{\ifcat a\noexpand#1\mathbf{#1}\else\bm{#1}\fi}
\newcommand{\unit}[1]{\ifcat a\noexpand#1\mathbf{\hat{#1}}\else\bm{\hat{#1}}\fi}
\makeatother

\providecommand{\mnras}{Mon. Not. R. Astron. Soc.}

\begin{document}
\begin{bibunit}[apsrev4-2]
\setbibunitnamespace{-main}

\title{Close Pulsar Pairs See Dark Matter Substructure through \\
the Gravitational-Wave Background}%

\date{\today}

\preprint{CALT-TH-2026-032}
\preprint{KA-TP-26-2026}
\preprint{N3AS-26-022}

\author{Abhiram Cherukupalli}
% \email{abhiram@caltech.edu}
\affiliation{California Institute of Technology,
1200 E. California Boulevard, Pasadena, CA 91125, USA}

\author{Vincent S. H. Lee}
% \email{vincentszehimlee@berkeley.edu}
\affiliation{Department of Physics, University of California,
Berkeley, CA 94720, USA}
\affiliation{Department of Physics, University of California, San Diego, La Jolla, CA 92093-0319, USA}

\author{Kim V. Berghaus}
% \email{kim.berghaus@kit.edu}
\affiliation{
Institute for Theoretical Physics (ITP), Karlsruhe Institute of Technology (KIT), Wolfgang-Gaede-Str. 1,
76131 Karlsruhe, Germany
}
\affiliation{
Institute for Astroparticle Physics (IAP), Karlsruhe Institute of Technology (KIT),
Hermann-von-Helmholtz-Platz 1, 76344 Eggenstein-Leopoldshafen, Germany}
%\affiliation{Walter Burke Institute for Theoretical Physics, California Institute of Technology,
%1200 E. California Boulevard, Pasadena, CA 91125, USA}

\author{Kathryn M. Zurek}
% \email{kzurek@caltech.edu}
\affiliation{Walter Burke Institute for Theoretical Physics, California Institute of Technology,
1200 E. California Boulevard, Pasadena, CA 91125, USA}

\begin{abstract}
Pulsar timing arrays detect signals from dark matter substructure, but a nanohertz gravitational-wave background (GWB) severely degrades their sensitivity. We show that timing pulsar pairs separated by sub-parsec distances avoids this degradation. The GWB imprints nearly the same perturbation on both pulsars and is suppressed in the difference of their residuals, whereas a nearby dark matter subhalo perturbs the two differently and survives. We find that an array with pulsar pairs separated by $0.1\,\mathrm{pc}$ improves the sensitivity to subhalos of order $\sim10^{-6}$ solar masses by more than an order of magnitude, relative to an otherwise identical array of widely separated pulsars. Dense globular clusters naturally host many such pairs, though the baryonic environment must be separated from a dark matter substructure signal.
\end{abstract}

\maketitle

\textbf{Motivation.}---
Observational evidence for dark matter (DM) spans galactic to cosmological scales, yet its fundamental nature remains one of the most pressing open questions in physics. Gravity is the only interaction DM is guaranteed to have. Imprints from the clustering of DM into 
sub-galactic scale structures therefore provide an irreducible window into its underlying microphysics. For instance, thermal relics such as weakly interacting massive particles do not cluster into subhalos with mass below $\sim 10^{-6}$ solar masses ($M_\odot$) \cite{Green:2005fa}, whereas axions can enhance galactic substructure on these scales~\cite{Hogan:1988mp, Kolb:1993zz, Zurek:2006sy, Buschmann:2019icd, Eggemeier:2019khm, Xiao:2021nkb}. 
Such low-mass subhalos are too light to host stars \cite{Bullock:2000wn, Benson:2001au, Bullock:2017xww, DES:2019ltu} and often too diffuse to effectively lens~\cite{Zurek:2006sy, 
Gilman:2019nap, Vegetti:2023mgp, 
Fairbairn:2017sil, VanTilburg:2026eym}, but their time-dependent gravitational potential may still be detectable. 

Precise timing measurements of 
radio pulses from millisecond pulsars (MSPs) are sensitive to small changes in the gravitational potential along the line of sight through Doppler and Shapiro effects.  They thereby probe 
the gravitational potential created by 
DM subhalos in the Earth-pulsar system
~\cite{Siegel:2007fz, Baghram:2011is, Clark:2015sha, Schutz:2016khr, Kashiyama:2018gsh, Dror:2019twh, Ramani:2020hdo, Lee:2020wfn, Lee:2021zqw, Gresham:2022biw, NANOGrav:2023hvm, Berghaus:2025kvn, Cherukupalli:2026cda, Foster:2026kfg}. 
And unlike lensing, MSPs are sensitive to relatively diffuse subhalos making them a powerful probe of the microscopic theory of DM~\cite{Ramani:2020hdo, Lee:2020wfn}. If the search for the gravitational signatures of DM subhalos were limited by white measurement noise alone, a next-generation pulsar timing array (PTA) enabled by the Square Kilometre Array (SKA) would constrain the abundance of DM subhalos in the $10^{-8}-10^2\, M_\odot$ mass range~\cite{Ramani:2020hdo, Lee:2020wfn, Lee:2021zqw}.

The timing data, however, now show evidence for a nanohertz stochastic gravitational-wave background (GWB)~\cite{NANOGrav:2023gor,EPTA:2023fyk, Reardon:2023gzh, Xu:2023wog, Miles:2024seg, InternationalPulsarTimingArray:2023mzf} potentially seeded by a cosmic population of supermassive-black-hole binaries (SMBHBs)~\cite{Phinney:2001di, NANOGrav:2023hfp, EPTA:2023xxk}. Although this evidence
is a triumph of the PTA science program, the GWB substantially degrades sensitivity to DM substructure
across the entire %
subhalo mass range, making the gravitational potentials induced by DM subhalos unreachable even with an advanced array~\cite{Lee:2021zqw,Foster:2026kfg,Cherukupalli:2026cda}. 

In this Letter, we propose %
separating a DM subhalo signal from the GWB
by exploiting their different spatial signatures, 
thereby restoring the sensitivity of PTAs to DM. At the nanohertz frequencies, $f\sim 1/T$, where the its power lies, the GWB is coherent over parsec scales, set by
\begin{equation}
\label{eqn:GWB_coherence_length}
\begin{aligned}
    d_{\rm coh}(f)
    &\equiv
    \frac{c}{2\pi f}\simeq
    1\,{\rm pc}
    \left(\frac{1.6\,{\rm nHz}}{f}\right) \sim \frac{c T}{2\pi},
\end{aligned}
\end{equation}
where $T \sim 20 \mbox { yr}$ is the observing time and 
$c$ is the speed of light. That is, pulsars separated by distances $d \ll d_\text{coh}$ see nearly the same perturbation from the GWB. By contrast, the gravitational influence of a DM subhalo is highly local, %
as it only depends on the proximity of a DM subhalo to a pulsar or its line of sight during the observation, characterized by a typical scale $r_\text{DM}$. %
One can therefore time a pair of pulsars separated by $r_\text{DM} \ll d\ll d_\text{coh}$
and consider the {\em difference}
of their timing residuals in which the GWB largely cancels %
but a subhalo signal persists.

This Letter proceeds as follows. We first briefly review the DM signal and the physical origin of the GWB coherence scale. We then construct the difference channel of a close pulsar pair, and use it to show how it can recover the DM sensitivity lost to the GWB.  Lastly, we discuss pulsar pairs in a globular cluster environment. %
Details and explicit derivations are left to the Supplemental Material (SM), 
to which we refer throughout the text.  %

\begin{figure}
    \centering
    \includegraphics[width=\columnwidth]{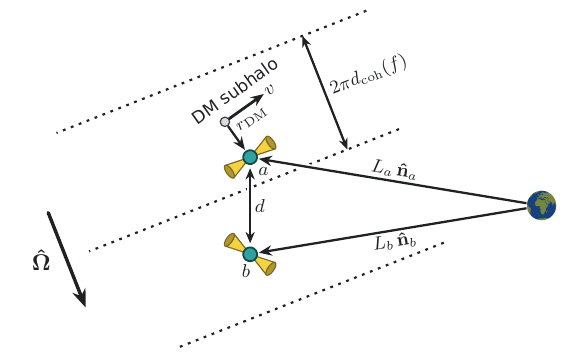}
    \caption{Two pulsars $a$ and $b$, located at $L_a\unit n_a$ and $L_b\unit n_b$ relative to Earth, with separation $d=\left|L_a\unit n_a-L_b\unit n_b\right|$. A plane-wave component of the GWB propagates along $\unit\Omega$ with wavelength $2\pi d_{\rm coh}(f)$. For $d\ll d_{\rm coh}(f)$, the GWB contributions to their timing residuals are nearly identical. By contrast, a loud DM subhalo moving with velocity $v$ passes close to only one of the pulsars, $r_\text{DM} \ll d$, and perturbs them differently.}
    \label{fig:schematic_strategy}
\end{figure}

\textbf{Signal from a DM flyby.}--- A transiting DM subhalo perturbs the observed arrival times of the pulses emitted from an MSP. This shift in the proper time is gauge-invariant by construction~\cite{Magi:2026upf, Lee:2026gzl}. In practice it is dominated by two contributions: a Doppler contribution for subhalos passing close to the pulsar, which accelerate it relative to the Earth, and a Shapiro contribution for subhalos passing close to the line of sight, which perturb the pulse's worldline~\cite{Dror:2019twh,Ramani:2020hdo}.\footnote{We neglect subhalos passing close to Earth as their signals are common across a close pulsar pair. %
}  We review the expression of the DM signal, which has been treated extensively in the literature, in SM Sec.~\ref{supp:sec:DM}. 

The strength of this perturbation is set by the subhalo mass and its distance to the Earth-pulsar system, which is determined by the density of DM subhalos $\rho_\text{DM}$, which, for simplicity, we assume to be composed of point-like subhalos of mass $M$ (for the generalization to diffuse subhalos, see Refs.~\cite{Ramani:2020hdo,Lee:2020wfn}). 
We approximate the
total signal from a population of these subhalos %
by the loudest flyby~\cite{Dror:2019twh, Ramani:2020hdo, Cherukupalli:2026cda}, which
will pass within a distance $r_{\rm DM}$ of a pulsar or its line of sight. The expression for $r_{\rm DM}$ simplifies in four different regimes of the DM flyby and is also reviewed in SM Sec.~\ref{supp:sec:DM}.  %

\textbf{Coherence scale of the GWB.}--- 
To understand how the GWB is correlated across pulsars, first consider the response of a pulsar to a single plane GW. As shown in Fig.~\ref{fig:schematic_strategy}, the pulsar $a$ is at a distance $L_a$ along $\unit n_a$ from the Earth, and a %
GW is propagating along $\unit\Omega$.
This wave induces a %
redshift $z_a\equiv\delta\nu_a/\nu_a$ in the frequency $\nu_a$ at which pulses arrive at the Earth. Working in the transverse-traceless gauge, $z_a$ reduces to a Shapiro contribution, 
which is 
the projected difference between the GW's metric perturbation at the pulse's reception event $x_E = (0, \mathbf{0})$, and at its emission event, $x_{P, a} = (-L_a/c, L_a\unit{n}_a)$~\cite{Romano:2023zhb} referred to as the \emph{Earth} (E) and \emph{pulsar} (P) terms, respectively. %
Fourier transforming in the arrival time, the redshift takes the form~\cite{Maggiore:2018sht}
\begin{equation} \label{eqn:z_response}
    z_a(f,\unit{\Omega})
    =
    \sum_{A=+,\times}
    F_a^A(\unit{\Omega})\,
    h_A(f,\unit{\Omega})
    \left(\underbrace{1}_{E}-\underbrace{e^{i\Phi_a}}_{P}\right),
\end{equation}
where $h_A(f,\unit{\Omega})$ is the Fourier amplitude of the wave's $A$ polarization at Earth and $F_a^A(\unit{\Omega})$ is the antenna pattern that encodes the projection onto $\unit{n}_a$, given explicitly in SM Sec.~\ref{supp:sec:ORF_expansion}. 
The Earth and pulsar terms sample the same wave at different points along its propagation, and so they differ only by a relative phase $\Phi_a(f,\unit{\Omega})={2\pi fL_a}(1+\unit{\Omega}\cdot\unit{n}_a)/c$.

The GWB is a stochastic superposition of many such GWs from sources distributed across the sky, all far enough away that their wavefronts are planar over the Earth--pulsar baseline. The correlation that this background induces between the timing residuals in pulsars $a$ and $b$ is encoded in its overlap reduction function (ORF) $\Gamma_{ab}(f)$. Taking it to be unpolarized and isotropic, the ORF is the sky average %
$\langle z_a z_b^*\rangle$~\cite{Romano:2023zhb},
\begin{equation} \label{eqn:ORF_explicit}
\begin{aligned}
    \Gamma_{ab}(f)
    =
    \frac{3}{2}
    &\sum_{A=+,\times}
    \int \frac{d^2\unit{\Omega}}{4\pi}\,
    F_a^A(\unit{\Omega})F_b^A(\unit{\Omega})
    \\
    &\quad\times
    \left(1-e^{i\Phi_a}\right)
    \left(1-e^{-i\Phi_b}\right),
\end{aligned}
\end{equation}
with the prefactor fixed by the normalization $\Gamma_{aa}\equiv1$. Any contribution whose phase varies rapidly with $\unit{\Omega}$ averages to zero, which occurs at length scales longer than $d_{\rm coh}$ defined in Eq.~\eqref{eqn:GWB_coherence_length}.

Expanding the product in Eq.~\eqref{eqn:ORF_explicit}, four terms contribute. The Earth--Earth term carries no $\unit{\Omega}$-dependent  phase and survives the sky average, giving the Hellings--Downs (HD) correlation~\cite{Hellings:1983fr}. %
The two Earth--pulsar cross terms carry a single phase, which varies rapidly with $\unit{\Omega}$ for pulsars at kiloparsec distances, far larger than $d_{\mathrm{coh}}(f)$, so they vanish upon averaging~\cite{Mingarelli_2018}.
The pulsar--pulsar term depends only on the difference $\Phi_a - \Phi_b$, and therefore on the distance between the pulsars rather than their individual distances from Earth. 

In a conventional PTA, the pulsars are separated from each other
by much more than $d_{\rm coh}(f)$, so the pulsar--pulsar term averages to zero as well. Each pulsar term cannot be predicted from the timing data of the others, and it sets an uncertainty floor for any analysis that exploits inter-pulsar correlations. The degraded sensitivity to DM found in Refs.~\cite{Foster:2026kfg, Cherukupalli:2026cda} therefore cannot be recovered even for signals correlated across the array, such as from subhalos passing near the Earth. %

If instead the pulsars are separated by a distance $d\ll d_{\rm coh}(f)$, on sub-parsec scales, their pulsar terms remain correlated~\cite{Mingarelli:2014xfa}. Taking all pulsars at the same distance from the Earth, $L_a=L_b \equiv L$,\footnote{Unequal pulsar distances are treated in SM Sec.~\ref{supp:dependence_orientation}. For pairs that are generically oriented relative to their line of sight, our conclusions are unchanged.}
the ORF to leading order in $d/d_\text{coh}$ has the form (see SM Sec.~\ref{supp:sec:ORF_expansion})
\begin{equation} \label{eqn:close_pulsar_ORF_schematic}
    \left|\Gamma_{ab}(f)\right|
    =
    \sqrt{\Gamma_{aa}\Gamma_{bb}}
    \left[
        1-%
        \frac{3}{40}\left(\frac{d}{d_\text{coh}(f)}\right)^2 %
    \right].
\end{equation}
As $\Gamma_{aa}=\Gamma_{bb}=1$, 
saturation of the Cauchy--Schwarz bound $|\Gamma_{ab}|\leq\sqrt{\Gamma_{aa}\Gamma_{bb}}$ would force the two GWB residuals to be identical. Therefore, up to corrections of size $\mathcal O\!\left[(d/d_{\rm coh}(f))^2\right]$, %
the GWB residual in one pulsar fixes that in the other.

\textbf{The close-pair difference channel.}--- Because the GWB perturbs the two pulsars nearly identically, its contribution largely cancels in the difference between their residuals. We now formalize this argument and demonstrate that differencing is optimal by analyzing the two timing residuals jointly. We model the residuals in pulsar $a$ as
\begin{equation}\label{eqn:residual_decomposition}
    \delta t_a(t) = %
    \sum_{i=0}^2 \xi_{a}^{(i)}  \phi_i (t) + \delta t_{\text{sig}, a}(t) + n_a (t),
\end{equation}
where $\phi_i$ are the basis functions of the quadratic %
pulsar timing model, with coefficients $\xi_a^{(i)}$, that encodes the uncertainty in the pulsar frequency and spin-down rate, and $\delta t_{\text{sig},a}$ is the deterministic signal we wish to detect. The noise $n_a$ is correlated between the two pulsars through its one-sided cross power spectral density (PSD) $S_{ab}(f)$, which we define in SM Sec.~\ref{supp:sec:correlations_noise_eigenbasis}. 
Here we consider two noise contributions.\footnote{We neglect additional pulsar-dependent red-noise contributions which degrade conventional array and close pair sensitivity alike.} First, finite timing precision leaves independent white noise in each residual, with one-sided PSD, $S_{\rm white}=2\Delta t t^2_\text{rms}$, set by the timing cadence $\Delta t$ and the root-mean-squared residual $t_\text{rms}$.
Second, the GWB contributes a PSD, described by its amplitude $A_{\rm GWB}$ and spectral index $\gamma$~\cite{NANOGrav:2023gor}, and is correlated between the two pulsars by $\Gamma_{ab}(f)$. We write the pulsar pair covariance as
$\mathbf{S}(f)=S_\text{white}\mathbf{s}(f)$, and $\mathbf{s}(f)$ is given by
\begin{equation} \label{eqn:GWB_pair_covariance}
    \mathbf{s}(f)
    =
    \mathbf{I}_2
    +
    \left(\frac{f_\star}{f}\right)^\gamma
    \begin{pmatrix}
        1 & \Gamma_{ab}(f) \\
        \Gamma_{ab}^*
        (f) & 1
    \end{pmatrix},
\end{equation}
where $f_\star \equiv f_{\rm yr}\left[{A_{\rm GWB}^2}/{(24\pi^2\Delta t\,t_{\rm rms}^2 f_{\rm yr}^3)}\right]^\frac{1}{\gamma}$ is the crossover frequency above which the noise transitions to white and $f_\text{yr}=1 \text{ yr}^{-1}$.

We estimate the sensitivity to a DM event by computing the signal-to-noise ratio (SNR). Diagonalizing the covariance matrix in Eq.~\eqref{eqn:GWB_pair_covariance} defines statistically independent \emph{channels}. Under our assumption $L_a=L_b$, $\Gamma_{ab}(f)$ is real, and these channels %
are simply the sum and difference of the residuals in the two pulsars, $\delta t_{\pm} =({\delta t}_{a} \pm {\delta t}_{b})/\sqrt{ 2 }$, with eigenvalues $s_{\pm}(f) = 1+\left[1\pm |\Gamma_{ab}(f)|\right]\left({f_\star}/{f}\right)^\gamma$. As discussed in SM Sec.~\ref{supp:sec:correlations_noise_eigenbasis}, the total squared SNR is then the sum of the squared SNRs in each channel,
\begin{equation}\label{eqn:pair_snr}
    \mathrm{SNR}^2
    =\frac{4}{S_\text{white}} \int_{1/T}^{\infty}\!df\,
    \sum_{\alpha = \pm}\frac{\big|\widetilde{\delta t}^\perp_{\text{sig},\alpha}(f)\big|^{2}}{s_{\alpha}(f)} \, ,
\end{equation}
where $\widetilde{\delta t}^\perp_{\text{sig},\alpha}$ is the part of the signal in channel $\alpha$ orthogonal to the timing model $\phi_i$, since any component parallel to it is degenerate with the fit and hence unobservable~\cite{Ramani:2020hdo, Lee:2020wfn,Cherukupalli:2026cda}. 

From Eq.~\eqref{eqn:close_pulsar_ORF_schematic}, we can see that if a DM signal survives in the difference channel, it would compete against a far weaker background, with a flatter spectrum, $\gamma \to \gamma -2$,
\begin{equation} \label{eqn:GWB_noise_sum_diff_channels}
\begin{aligned}
    s_-(f)
    &\simeq 
    1
    +
    \left(\frac{f_\star^{(-)}}{f}\right)^{\gamma-2},
\end{aligned}
\end{equation}
and a crossover to white noise that happens at a much lower frequency, $f_\star^{(-)}\ll f_\star$, 
\begin{equation}\label{eqn:GWB_difference_crossover}
  f_{\star}^{(-)}(d)\sim f_\star\left[{\frac{3}{40}}\left(\frac{d}{d_{\rm coh}(f_\star)}\right)^2\right]^\frac{1}{\gamma-2} .
\end{equation}
The largest gains are thus at the low frequencies where the GWB would otherwise dominate.\footnote{The results from Ref.~\cite{Cherukupalli:2026cda}, which provide analytical scalings for the DM signal reach
in a channel with noise of the form $s(f)=1+(f_\star/f)^\gamma$, can be directly extended to the difference channel. %
}~As $d$ decreases below a separation $d_\text{white}$, the residual GWB drops below the 
white noise at all observable frequencies,  
defined by $f_\star^{(-)}(d_{\rm white})T = 1$.

\textbf{Improved sensitivity by timing close pairs.}--- We now turn to presenting our main results.  Throughout, we assume that each pulsar is timed for $T = 20\, $yr with $\Delta t=2\,\mathrm{wk}$ and $t_{\rm rms}=50\,\mathrm{ns}$. We fix the spectral index to the %
predicted %
SMBHB value, $\gamma=13/3$~\cite{Phinney:2001di}, and adopt the corresponding best-fit amplitude $A_{\rm GWB}=2.4\times10^{-15}$ from Ref.~\cite{NANOGrav:2023gor}. At each subhalo mass $M$, we model the subhalo population as Poisson distributed with 
number density $\bar{n}=\rho_{\rm DM}/M$ and isotropic speed $v=340\, \mathrm{km/s} \sim 10^{-3} c$. 

\begin{figure}
    \centering
    \includegraphics[width=\linewidth]{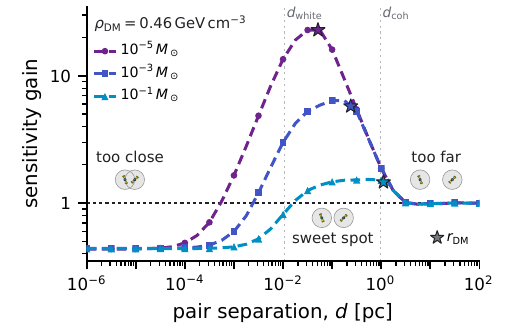}
    \caption{%
    The median SNR (normalized to $d\to \infty$) of the loudest subhalo flyby near a single pulsar pair as a function of the pair separation $d$, for subhalo masses $M=10^{-5},\,10^{-3},\,10^{-1}\,M_\odot$. Also indicated are the three characteristic separations governing the difference channel: $d_\text{coh}$ and the mass-dependent $r_\text{DM}$ (shown as a star), which respectively mark the onsets of GWB and DM suppression, and $d_\text{white}$, below which the channel is white-noise dominated. }
\label{fig:sensitivity_gain}
\end{figure}

In Fig.~\ref{fig:sensitivity_gain}, fixing $\rho_\text{DM}=0.46 \, \mathrm{GeV\, cm^{-3}}$, we plot the median SNR of the loudest subhalo flyby near a single close pulsar pair, relative to a well-separated pair, as a function of the pair separation $d$. For simplicity, we first consider only flybys near the pulsars whose signals are Doppler-dominated. 
For $d > d_{\rm coh}$, nothing is gained. Below $d_{\rm coh}$, the residual GWB in the difference channel is suppressed as $(d/d_{\rm coh})^2$ and the gain sets in. To exploit this gain, we need to time pairs in the window $r_\text{DM}\ll d\ll d_\text{coh}$. This window exists only at small subhalo masses, where the number density is high enough; this is not the case at $M \sim 10^{-1}M_\odot$. There, we are instead in the regime $d \lesssim r_\text{DM}$, where the two pulsars also experience similar DM signals,
leaving only their tidal residual in the difference channel, with power suppressed as $(d/r_\text{DM})^2$. Both the DM signal and the GWB are suppressed as $d^2$, so in the window $d_\text{white} \ll d \ll r_\text{DM}$, the sensitivity depends only weakly on $d$.  Below both scales, $d \ll\min(r_\text{DM}, d_\text{white})$, the DM signal falls relative to the now fixed white-noise-dominated difference channel and the gain drops sharply. The sensitivity is then set by the sum channel, whose sensitivity is order-unity worse than the well-separated case.

An array of many pulsar pairs offers far more opportunities for a close subhalo flyby, so the loudest event across the array has a much larger SNR (and a much smaller $r_\text{DM}$) than in the single-pair case. This motivates us to consider an %
array of $100$ %
pulsar pairs ($200$ pulsars), each with separation $d$ and located at a common distance $L = 5$~kpc from Earth, with each pulsar timed to our benchmark. We define the reach as the smallest $\rho_\text{DM}$ for which the loudest subhalo across the array has $\mathrm{SNR}\geq4$ in $90\%$ of realizations~\cite{Dror:2019twh}. We sample all relevant subhalos passing close to a pulsar or its line of sight %
and, as a conservative approximation, assume that distinct pairs are well-separated from each other and neglect the order-unity gain from exploiting the GWB correlations between %
pairs (including such correlations for pairs that are instead close to each other further suppresses the GWB, as discussed below). %
Together, these approximations allow us to treat each pair independently; numerical details are provided in SM Sec.~\ref{supp:numerics}.

\begin{figure*}
    \centering
    \includegraphics[width=\linewidth]{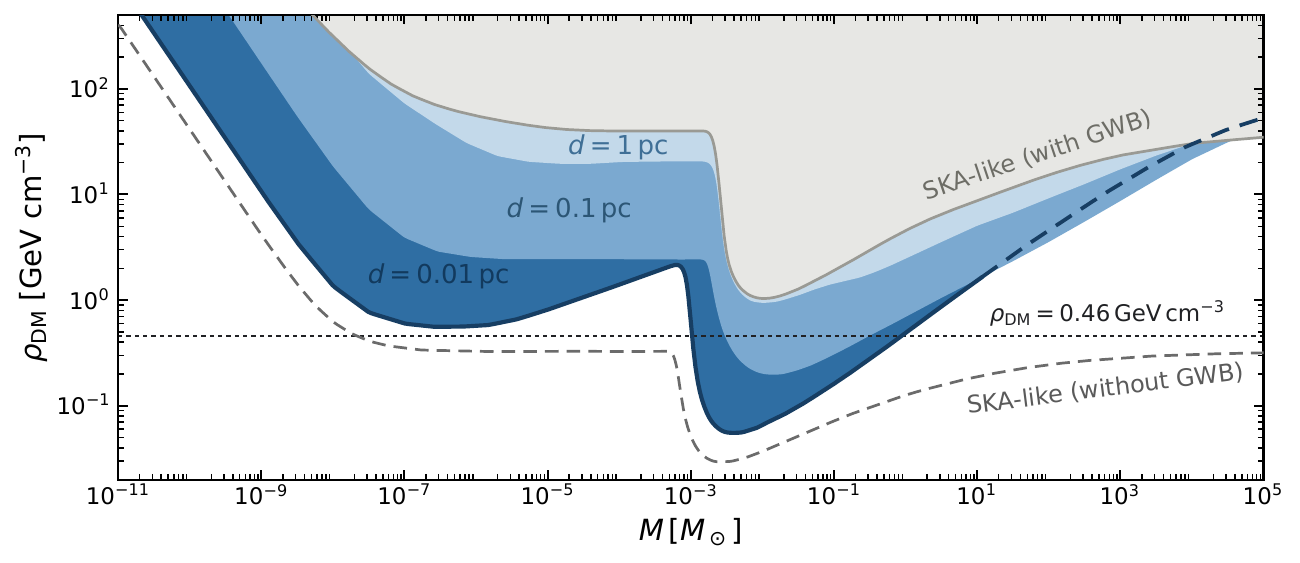}
    \caption{
    The minimum DM density in substructure $\rho_\text{DM}$ for which
    $90\%$ of random draws of the subhalo population contain at least one %
    subhalo with $\mathrm{SNR}\geq 4$ versus subhalo mass, $M$. Shown in blue is the reach for the close-pair array described in the main text, with $100$ pulsar pairs separated by $d=0.01$, $0.1$, and $1\,\mathrm{pc}$. Shown in gray is the reach in a conventional SKA-like array with and without a GWB from Ref.~\cite{Cherukupalli:2026cda}. Both arrays are timed to the same benchmark. The reach is Doppler-dominated to the left of $M\sim10^{-3}M_\odot$ and Shapiro-dominated to the right.
    }
    \label{fig:money_plot}
\end{figure*}

In Fig.~\ref{fig:money_plot}, we present the resulting reach across a mass range of $10^{-11}\,M_{\odot}-10^5\,M_{\odot}$ for pair separations $d=0.01$, $0.1$, and $1\,\mathrm{pc}$. For comparison, we show the reach with and without a GWB in a conventional $200$-pulsar array from Ref.~\cite{Cherukupalli:2026cda}, which uses the same timing benchmark. The gap between these reference curves quantifies the DM sensitivity lost to the GWB. The reach is dominated by the Doppler contribution for $M\lesssim10^{-3}M_\odot$, and by Shapiro for $M\gtrsim10^{-3}M_\odot$. At the lower-mass end of both the Doppler- and Shapiro-dominated ranges, the loudest flybys have a small enough $r_\text{DM}$ %
that a $d=0.01\,\mathrm{pc}$ pair suppresses the GWB without strongly suppressing the DM signal, recovering nearly all the sensitivity lost to the GWB. At higher masses, $r_\text{DM}$ eventually exceeds $0.01\,\mathrm{pc}$ and less sensitivity is recovered. The analytical estimate of this reach in the limit $r_{\rm DM}\gg vT$, presented in SM Sec.~\ref{supp:dop_static}, agrees well with the numerics. SM Sec.~\ref{supp:continuum_angular_structure} provides a complementary interpretation of the close-pair difference channel as a high-pass filter in angular space.

\textbf{Outlook.}--- 
Dense globular-cluster cores naturally host such close pairs~\cite{Mingarelli:2014xfa,Ng_2022, SKAPulsarScienceWorkingGroup:2025cun}. Terzan~5, for example, hosts $48$ 
localized MSPs concentrated in a core of $\sim0.3\,\mathrm{pc}$
~\cite{Padmanabh_2024}, 
and there may be at least $200$ more radio-visible MSPs yet to be discovered~\cite{urquhart2026newdeepradiocontinuum}. %
If these undiscovered pulsars follow the same observed %
distribution, Terzan~5 alone could contain
$\sim70$ %
pairs separated by less than $0.1\,\mathrm{pc}$ (SM Sec.~\ref{supp:potential_pairs}). %
Other globular clusters, such as 47~Tucanae~\cite{Freire_2001, Chen:2026uwh}, also host timed MSPs and their cores would contribute additional pairs. %
Given the large number of pulsars within 1~pc, the GWB %
would be correlated not just in pairs but across multiple pulsars, which allows us to construct channels that nearly eliminate the GWB (SM Sec.~\ref{supp:GWB_joint_suppress}). %

There are three questions, however, that will ultimately decide the feasibility of detecting DM substructure using MSPs in globular clusters. First, can they be timed precisely enough? They are faint, making our $t_\mathrm{rms}=50\,\mathrm{ns}$ benchmark demanding. However, even if they can only be timed to sub-$\mu$s precision, the close-pair advantage can persist (SM Sec.~\ref{supp:degrading_timing}); though, pulsar-intrinsic red noise may still be a challenge. %
Second, how much DM substructure is there to find in globular clusters? At present, there is no convincing evidence that globular clusters bind DM subhalos to their gravitational potential~\cite{Reynoso_Cordova_2022,Garani:2023esk}. However, clusters that are closer to the Galactic center, such as Terzan~5, have an ambient DM density that is substantially higher than the local value, though tidal disruption of DM subhalos in the inner Galaxy must be taken into account~\cite{Lee:2020wfn}.  Third, can the DM substructure signal be separated from the baryonic background? Globular cluster cores are baryon dense. For instance, Terzan~5's core has a mass density $\rho_\text{baryon}\sim 6\times 10^{7} \, \mathrm{GeV/cm^3}$~\cite{Prager:2016puh}, and baryonic substructure flybys can mimic DM. This background is not completely degenerate with DM substructure, since %
most of the mass is in stars moving at the cluster's virial speed, roughly $v\sim10^{-4}\, c$ for Terzan~5, which is an order of magnitude slower than the ambient DM. As a result, we expect stellar flybys to vary slowly over the observing baseline and may be absorbed by 
additional terms in the timing model. On the other hand, asteroid-to-planetary-scale baryonic substructure, whose abundance is unknown, likely cannot be removed this way.  
A dedicated %
study of these questions is left to future work.

Next-generation PTAs will place the 
GWB on firm observational footing.
The PTA program is, however, far richer than gravitational-wave detection~\cite{NANOGrav:2023hvm}, and pulsar timing is a promising
probe of low-mass DM substructure. For this DM %
search, %
the GWB is not a signal but a background. We have shown that differencing the residuals of close pulsar pairs removes this common background while preserving a DM substructure signal, establishing close-pair differencing as a detection technique that is not limited by the GWB.

\textbf{Acknowledgements.}---  A.C. was supported by the Margaret Leong and Michael P. Checca Summer Undergraduate Research Fellowship. V.L. is supported by the Network for Neutrinos, Nuclear Astrophysics and Symmetries (N3AS) through the National Science Foundation Physics Frontier Center, Grant No. PHY-2020275. K.B. is funded by the Deutsche Forschungsgemeinschaft (DFG, German Research Foundation) through the Emmy Noether Programme Project No.~548044346.  K.Z. is supported by the U.S. Department of Energy, Office of Science, Office of High Energy Physics, under Award No. DE-SC0011632, the Walter Burke Institute for Theoretical
Physics, the Heising-Simons Foundation, and a Simons Investigator award. The authors acknowledge the use of %
OpenAI's Codex and Anthropic's Claude Code as supplementary research-assistance tools for implementing numerics, making figures, finding mathematical tricks, and editing prose. The scientific ideas, codebase, derivations, interpretation of the results, and manuscript writing are the work of the authors.

\let\savedaddcontentsline\addcontentsline
\renewcommand{\addcontentsline}[3]{}
\putbib[biblio]
\let\addcontentsline\savedaddcontentsline
\end{bibunit}

\let\savedaddcontentsline\addcontentsline
\renewcommand{\addcontentsline}[3]{}
\let\addcontentsline\savedaddcontentsline

\addtocontents{toc}{\protect\setcounter{tocdepth}{1}}

\clearpage
\onecolumngrid
\begin{bibunit}[apsrev4-2]
\setbibunitnamespace{-supp}

\begin{center}
\textbf{\large SUPPLEMENTARY MATERIAL}\\[0.2cm]
\textbf{``Close Pulsar Pairs See Dark Matter Substructure through \\
the Gravitational-Wave Background"}\\[0.2cm]
Abhiram Cherukupalli, Vincent S. H. Lee, Kim V. Berghaus, and Kathryn M. Zurek
\end{center}

\setcounter{section}{0}
\setcounter{equation}{0}
\setcounter{figure}{0}
\setcounter{table}{0}

\setcounter{secnumdepth}{3}
\setcounter{tocdepth}{1}

\renewcommand{\thesection}{S\arabic{section}}
\renewcommand{\thesubsection}{\thesection.\Alph{subsection}}
\renewcommand{\theequation}{S\arabic{equation}}
\renewcommand{\thefigure}{S\arabic{figure}}
\renewcommand{\thetable}{S\arabic{table}}

\makeatletter
\onecolumn@grid@setup
\let\set@footnotewidth\set@footnotewidth@one
\renewcommand{\p@subsection}{}
\makeatother

\tableofcontents

\section{The dark matter signal}
\label{supp:sec:DM}

In this section, we review the dark matter (DM) signal and the typical encounter distance scale $r_\text{DM}$ of the loudest flyby. Both have been extensively developed in the literature~\cite{Dror:2019twh,Ramani:2020hdo,Lee:2020wfn,Lee:2021zqw,NANOGrav:2023hvm,Foster:2026kfg,Cherukupalli:2026cda}. Below, we summarize the relevant results and 
discuss how they are applied in our analysis. %

A DM subhalo transiting near the Earth--pulsar system shifts the observed arrival time of the pulses arriving from the pulsar. This gauge-invariant time shift can be decomposed into gauge-dependent Doppler ($\mathcal{D}$), Shapiro ($\mathcal{S}$) and Einstein ($\mathcal{E}$) contributions~\cite{Magi:2026upf,Lee:2026gzl}
\begin{equation}
\label{eqn:delta_t_t}
	\delta t_{\mathrm{DM}}(t) =\,  \delta t_{\mathcal{D}}^{(E)}(t) - \delta t_{\mathcal{D}}^{(P)}(t) + \delta t_{\mathcal{S}}(t) + \delta t_{\mathcal{E}}^{(E)}(t) - \delta t_{\mathcal{E}}^{(P)}(t) \,  ,
\end{equation}
where the $(E)$ and $(P)$ superscripts correspond to the Earth and pulsar terms, respectively. In Ref.~\cite{Cherukupalli:2026cda}, we evaluated these terms by working in the Newtonian gauge and list the resulting expressions here for completeness. Here $\vec{r}_E$ is the location of the Earth and  $\vec{r}_P =\vec{r}_E + L\unit{n}$ the location of the pulsar at a distance $L$ along $\unit{n}$ away from the Earth. We introduce three different parametrizations of the subhalo's trajectory
\begin{align}\label{eqn:DM_parameterization}
	\vec{x}_{\mathrm{DM}} &= \vec{r}_E+ \vec{b}_E + \vec{v}(t-t_E) \nonumber \\
	&= \vec{r}_P+\vec{b}_P + \vec{v}(t-t_P) \nonumber \\
	&= \frac{\vec{r}_E+\vec{r}_P}{2} + \left[b_{\parallel}+v_{\parallel}(t-t_{\perp})\right]\unit{n} + \left[\vec{b}_{\perp} + \vec{v}_{\perp}(t-t_{\perp})\right] \, ,
\end{align}
where $\vec{b}_E$, $\vec{b}_P$ and $\vec{b}_{\perp}$ are the impact parameters relative to Earth, the pulsar, and the line of sight, $t_E$, $t_P$ and $t_{\perp}$ are the corresponding times of closest approach. $b_\parallel$ is the DM’s longitudinal distance from Earth along the Earth-pulsar line of sight, $v_{\parallel}=\vec{v}\cdot\unit{n}$ and $\vec{v}_{\perp}=\vec{v} - v_{\parallel}\unit{n}$ are the parallel and orthogonal components of $\vec{v}$ relative to $\unit{n}$. The Doppler contribution, dominant for subhalos passing close to the pulsar or the Earth, then takes the form,
\begin{subequations}\label{eqn:DM_evaluated}
\begin{align}
\delta t_{\mathcal D}^{(E)}(t)
&=
\frac{GM}{v^2c}
\left[
\sqrt{1+x_E^2}\,(\unit{n}\cdot \unit{b}_E)
-\sinh^{-1}(x_E)(\unit{n}\cdot \unit{v})
\right],
\label{eqn:DM_evaluated_doppler_earth}
\\
\delta t_{\mathcal D}^{(P)}(t)
&=
\frac{GM}{v^2c}
\left[
\sqrt{1+x_P^2}\,(\unit{n}\cdot \unit{b}_P)
-\sinh^{-1}(x_P)(\unit{n}\cdot \unit{v})
\right],
\label{eqn:DM_evaluated_doppler_pulsar}
\end{align}
\end{subequations}
where we have defined dimensionless time variables, $x_E\equiv v(t-t_E)/b_E$ and $x_P\equiv v(t-t_{P}-L/c
)/b_P$. The Einstein contribution, relevant for the same subhalos but subdominant by $v/c$ relative to the Doppler, is given by
\begin{subequations}\label{eqn:DM_evaluated_einstein}
\begin{align}
    \delta t_{\mathcal E}^{(E)}(t)
&=
-\frac{GM}{vc^2}\sinh^{-1}(x_E),
\label{eqn:DM_evaluated_einstein_earth}
\\
\delta t_{\mathcal E}^{(P)}(t)
&=
-\frac{GM}{vc^2}\sinh^{-1}(x_P).
\label{eqn:DM_evaluated_einstein_pulsar}
\end{align}
\end{subequations}
Finally, the Shapiro contribution, relevant for subhalos passing close to the line of sight, takes the form
\begin{align}\label{eqn:DM_Shapiro_2}
	\delta t_{\mathcal{S}}(t) &= \frac{2GM}{c^3}\log\left[\frac{r_{\parallel}+(L/2)+\sqrt{r_{\perp}^2+\left[r_{\parallel}+(L/2)\right]^2}}{r_{\parallel}-(L/2)+\sqrt{r_{\perp}^2+\left[r_{\parallel}-(L/2)\right]^2}}\right],
\end{align}
where %
\begin{align}\label{eqn:r_par_def}
	r_{\parallel} &\equiv b_{\parallel}+v_{\parallel}\left(t-L/(2c)-b_\parallel/c-t_\perp\right),
    \\
    \label{eqn:r_perp_def}
	r_{\perp}^2&\equiv b_{\perp}^2 + v_{\perp}^2\left(t-L/(2c)-b_\parallel/c-t_\perp\right)^2.
\end{align}
This is suppressed by $(v/c)^2$ %
relative to the Doppler contribution. However, subhalos can get much closer to the line of sight than to the pulsar or the Earth, and thus the Shapiro contribution can still be comparable to the Doppler terms.%

\begin{figure}
    \centering
    \includegraphics[width=0.8\linewidth]{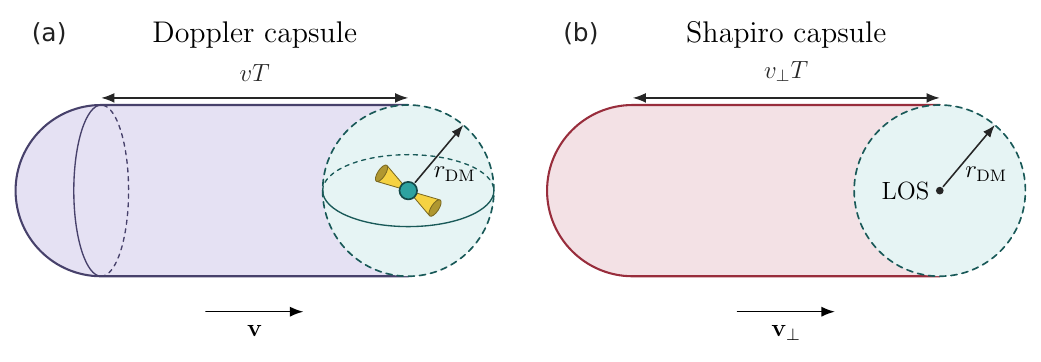}
    \caption{The volumes of all initial positions from which a subhalo comes within $r_\text{DM}$ of a pulsar or its line of sight during the observation window $T$, for a fixed direction of the subhalo velocity $\vec{v}$, form (a) the Doppler capsule and (b) the Shapiro capsule, respectively.%
    }
    \label{fig:dm_capsules}
\end{figure}

We now discuss the typical DM subhalo encounter distance scale, which we denote as $r_\text{DM}$. For our benchmark of $v=340\, \mathrm{km/s}\sim 10^{-3} c$ and $T=20\, \mathrm{yr}$, a subhalo moves a distance $v T \simeq 7\times10^{-3}\, \mathrm{pc}$ over the observing duration. Therefore, a subhalo initially farther than $r_{\mathrm{DM}}$ can come within a distance of $r_{\mathrm{DM}}$ during the observation window. For a fixed direction of $\mathbf{v}$, the initial positions from which this occurs fill a capsule of radius $r_\text{DM}$ and length $vT$. We illustrate these geometries in Fig.~\ref{fig:dm_capsules}. In Ref.~\cite{Cherukupalli:2026cda}, we refer to these as the ``Doppler capsule" for subhalos near the pulsar, with volume $V_\mathcal{D}$, and the ``Shapiro capsule" for subhalos near the line of sight, with volume $V_\mathcal{S}$. The Doppler capsule consists of a sphere with a radius $r_{\mathrm{DM}}$ centered at the pulsar, and a cylinder with radius $r_{\mathrm{DM}}$ and length $vT$. The Shapiro capsule consists of a cylinder with radius $r_{\mathrm{DM}}$ and length $L$, and a cuboid with width $v_{\perp}T$, height $2r_{\mathrm{DM}}$, and depth $L$. Their volumes are thus given by
\begin{equation} \label{eqn:capsule_volumes}
\begin{split}
    V_\mathcal{D} (r_\text{DM}) &= \frac{4\pi}{3}r_\text{DM}^3 +\pi r_\text{DM}^2vT, \\
    V_\mathcal{S} (r_\text{DM}) &= \pi r_\text{DM}^2 L + 2r_{\text{DM}} L v_\perp T \, ,
\end{split}
\end{equation}
where we replace $v_\perp$, whose value changes with the direction of $\unit{v}$, with its mean %
$\bar{v}_\perp %
\simeq 270 \, \mathrm{km/s}$.

We define the encounter distance scale $r_\text{DM}$ such that, in half of all realizations, at least one subhalo comes within $r_\text{DM}$ of a pulsar in the array or its line of sight during the observation. The relevant distance is measured to the pulsar when the Doppler effect dominates the signal, and to the line of sight when the Shapiro effect dominates.
The number of subhalos that pass within $r_\text{DM}$ during the observation window has mean $\bar{n} V(r_\text{DM})N_{P}$, with $V$ the capsule volume given in Eq.~\eqref{eqn:capsule_volumes}, and $N_{P}$ the number of %
pulsars %
in the array. Since the subhalos are Poisson distributed, requiring that at least one such subhalo exists in half of the realizations gives $1 - e^{-\bar{n} V(r_\text{DM})N_{P}} = 1/2$, which implies $N_{P}V(r_\text{DM}) = M\ln 2/\rho_\text{DM}$.
On the other hand, the volumes in Eq.~\eqref{eqn:capsule_volumes} take different expressions in two different limits: $r_{\mathrm{DM}}\gg vT$ or $r_{\mathrm{DM}}\ll vT$, known as the static or dynamic limits, corresponding to whether the subhalo moves substantially during the observing time $T$~\cite{Dror:2019twh}. In the cases where the Doppler effect dominates the signal, we find 
\begin{equation} \label{eqn:Doppler_rDM}
\begin{gathered}
    \text{\bfseries Doppler-dominated flybys}\\[0.5em]
    r_\text{DM} \approx \left\{
    \begin{array}{l@{\;}c@{\;}l@{\quad}l@{\quad}l}
        \displaystyle\left(\frac{3M\ln 2}{4\pi \rho_\text{DM}N_{P}}\right)^{1/3}
        & \simeq &
        \displaystyle 0.2\,\mathrm{pc}
        \left(\frac{M}{10^{-3}\,M_\odot}\right)^{1/3}\left(\frac{1}{N_{P}}\right)^{1/3}
        \left(\frac{0.46\,\mathrm{GeV/cm^3}}{\rho_\text{DM}}\right)^{1/3},
        & r_\text{DM}\gg vT
        & (\text{static limit})
        \\[1.5em]
        \displaystyle\sqrt{\frac{M\ln 2}{\pi \rho_\text{DM}N_{P}vT}}
        & \simeq &
        \displaystyle 2\times10^{-3}\,\mathrm{pc}
        \left(\frac{M}{10^{-9}\,M_\odot}\right)^{1/2}
        \left(\frac{1}{N_{P}}\right)^{1/2}\left(\frac{0.46\,\mathrm{GeV/cm^3}}{\rho_\text{DM}}\right)^{1/2},
        & r_\text{DM}\ll vT
        & (\text{dynamic limit}).
    \end{array}
    \right.
\end{gathered}
\end{equation}
Similarly, with our $L=5\, \mathrm{kpc}$ benchmark, if the Shapiro delay dominates the signal, then %
\begin{equation}\label{eqn:Shapiro_rDM}
\begin{gathered}
    \text{\bfseries Shapiro- dominated flybys}\\[0.5em]
    r_\text{DM} \approx \left\{
    \begin{array}{l@{\;}c@{\;}l@{\quad}l@{\quad}l}
        \displaystyle\sqrt{\frac{M\ln 2}{\pi \rho_\text{DM}N_{P} L}}
        & \simeq &
        \displaystyle 0.2 \, \mathrm{pc}
        \left(\frac{M}{10\,M_\odot}\right)^{1/2}
        \left(\frac{1}{N_{P}}\right)^{1/2}\left(\frac{0.46\,\mathrm{GeV/cm^3}}{\rho_\text{DM}}\right)^{1/2},
        & r_\text{DM}\gg \bar{v}_\perp T
        & (\text{static limit})
        \\[1.5em]
        \displaystyle\frac{M\ln 2}{2 \rho_\text{DM}N_{P} L\bar{v}_\perp T}
        & \simeq &
        \displaystyle 10^{-3}\,\mathrm{pc}
        \left(\frac{M}{10^{-3}\,M_\odot}\right)
        \left(\frac{1}{N_{P}}\right)\left(\frac{0.46\,\mathrm{GeV/cm^3}}{\rho_\text{DM}}\right),
        & r_\text{DM}\ll \bar{v}_\perp T
        & (\text{dynamic limit})\, .
    \end{array}
    \right.
\end{gathered}
\end{equation}
The expressions above show that at low enough masses, our desired window, $r_\text{DM} \ll d \ll d_\text{coh}$, exists, where the GWB contribution cancels but the DM signal does not. For an array with $100$ %
pulsar pairs, as considered in the main text, $r_\text{DM}$ is even smaller, as shown in Eqs.~\eqref{eqn:Doppler_rDM}--\eqref{eqn:Shapiro_rDM}, %
and our desired window further widens relative to the single pair case. Equations.~\eqref{eqn:Doppler_rDM}-\eqref{eqn:Shapiro_rDM} assume disjoint capsules. However, overlap between the two capsules in a pair reduces the total volume. Neglecting overlaps between distinct pairs, this volume lies between $N_{P}V(r_\text{DM})/2$ and $N_PV(r_\text{DM})$, corresponding to fully overlapping and disjoint capsules in a pair.\footnote{We note here that the $r_{\mathrm{DM}}$ defined as above, which depends on $\rho_{\mathrm{DM}}$, is not the same quantity as the ``detection distance" considered in Ref.~\cite{Cherukupalli:2026cda}, which is defined as the smallest distance a flyby needs to get to generate a signal with $\text{SNR}=4$.}

\section{The GWB overlap reduction function}

The correlation structure of the pulsar timing measurements between pulsars is encoded in a quantity known as the overlap reduction function (ORF), defined as the cross-correlation of the timing residuals of a pulsar pair and normalized to the autocorrelation of a single pulsar.%
In particular, the ORF of an isotropic GWB was derived in Ref.~\cite{Hellings:1983fr} under two conditions: each pulsar is far from the Earth, and any two distinct pulsars are far from each other, both compared to the GW wavelength. %
This form of the ORF is widely used by PTA collaborations in searches for the GWB (see Ref.~\cite{Taylor:2021yjx} for a review). Corrections to this form for pulsar pairs at small separations were
computed numerically in Ref.~\cite{Mingarelli:2014xfa}.  

In this section, we rederive the ORF of the GWB, where we keep the first condition but drop the second, \textit{i.e.}, we allow the separation between two pulsars to be smaller than the GW wavelength. As we will show below, the pulsar--pulsar term then survives and drives the ORF to unity as $d/d_\text{coh}(f)\to0$, with corrections of order $(d/d_\text{coh}(f))^2$.

\subsection{Large and small pulsar separations}
\label{supp:sec:ORF_expansion}

The ORF of the GWB can be written in the following form~\cite{Romano:2023zhb}
\begin{equation} \label{eqn:GWB_ORF_exact}
    \Gamma^\text{GWB}_{ab}(f)
    =
    \frac{3}{2}\int \frac{d^2\Omega}{4\pi}
    \sum_{A=+,\times}
    F_a^A(\unit{\Omega})\,
    F_b^A(\unit{\Omega})
    \left[\underbrace{1}_{E}-\underbrace{e^{i\Phi_a(f,\unit{\Omega})}}_{P_a}\right]
    \left[\underbrace{1}_{E}-\underbrace{e^{-i\Phi_b(f,\unit{\Omega})}}_{P_b}\right],
\end{equation}
where $\Phi_a(f,\unit{\Omega})={2\pi fL_a}(1+\unit{\Omega}\cdot\unit{n}_a)/c$, $\unit{\Omega}$ is the propagation direction of the gravitational wave, $L_a$ the distance to pulsar $a$, similarly for pulsar $b$ and %
\begin{equation} \label{eqn:antenna_pattern}
    F_a^A(\unit{\Omega}) = \frac{\unit{n}_a^i \unit{n}_a^j}{2(1+\unit{\Omega}\cdot \unit{n}_a)} e_{ij}^A (\unit{\Omega}),
\end{equation}
is the antenna pattern, with $e_{ij}^A(\unit{\Omega})$ being the polarization tensor. In each bracket in Eq.~\eqref{eqn:GWB_ORF_exact}, the 1 corresponds to the Earth term ($E$), common to all pulsars, and the term proportional to $e^{i\Phi_a}$ is the pulsar term ($P_a$), which depends on the distance to pulsar $a$. Expanding the product gives an Earth--Earth term ($EE)$, a pulsar--pulsar term ($PP$), and two Earth--pulsar cross terms ($EP$).

We first impose the condition that the pulsars are far from the Earth, $L_a, L_b \gg d_{\rm coh}(f) = c/(2\pi f)$.
In this limit, the relative phase $\Phi_a$ rapidly oscillates with $\unit{\Omega}$, so the cross-correlation terms ($EP$) between the Earth and pulsar are suppressed, and the ORF takes the form~\cite{Mingarelli_2018},

\begin{equation} \label{eqn:GWB_ORF_short_wavelength}
    \Gamma^\text{GWB}_{ab}
    \approx
    \frac{3}{2}\int \frac{d^2\Omega}{4\pi}
    \sum_{A=+,\times}
    F_a^A(\unit{\Omega})\,
    F_b^A(\unit{\Omega})
    \left[\underbrace{1}_{EE}+\underbrace{e^{i\Delta\Phi_{ab}(f,\unit{\Omega})}}_{P P}\right],
\end{equation}
where $\Delta \Phi_{ab} (f, \unit{\Omega})\equiv \Phi_a - \Phi_b= 2\pi f[(L_a-L_b)+\unit{\Omega}\cdot \vec{d}]/c$, and we denote the separation between the pulsars as $\vec{d}=L_a \unit{n}_a-L_b \unit{n}_b$. 

For well-separated pulsars, $d\gg d_\text{coh}(f)$, $\Delta \Phi_{ab} (f, \unit{\Omega})$ also oscillates rapidly with $\unit{\Omega}$ and the pulsar term correlation between distinct pulsars ($PP$) is suppressed. For the same pulsar $a=b$, by contrast, $\Delta\Phi_{aa}= 0$ and we have
\begin{equation}
    \Gamma^\text{GWB}_{ab} \approx \left[1+\delta_{ab}\right]\times \frac{3}{2}\int \frac{d^2\Omega}{4\pi } \sum_{A=+, \times } F_a^A (\unit{\Omega}) F_b^A (\unit{\Omega}).  %
\end{equation}
This integral produces the familiar Hellings--Downs (HD) correlation~\cite{Hellings:1983fr},
\begin{equation}\label{eqn:HD}
    \Gamma^\text{HD}_{ab}\equiv \frac{3}{2}\int \frac{d^2\Omega}{4\pi } \sum_{A=+, \times } F_a^A (\unit{\Omega}) F_b^A (\unit{\Omega}) = \frac{3}{2}x_{ab}\log(x_{ab})-\frac{1}{4}x_{ab}+\frac{1}{2} \, ,
\end{equation}
where $x_{ab}\equiv (1-\cos \zeta_{ab})/2$, and $\zeta_{ab}\equiv \cos^{-1}(\unit{n}_a\cdot \unit{n}_b)$ is the angular separation between pulsars $a$ and $b$. Since $\Gamma_{aa}^\text{HD}=1/2$ in Eq.~\eqref{eqn:HD}, the ORF simplifies to~\cite{Romano:2023zhb}
\begin{equation} \label{eqn:GWB_ORF_well_separated}
    \Gamma^\text{GWB}_{ab}= \Gamma^\text{HD}_{ab} +\Gamma^\text{HD}_{aa} \delta_{ab}=\Gamma^\text{HD}_{ab}+\frac{1}{2}\delta_{ab} \, ,
\end{equation}
where we see that $\Gamma^\text{GWB}_{aa}=1$ is properly normalized. The corrections from incompletely decohered Earth--pulsar cross terms ($EP$) are of order $d_\text{coh}(f)/L\lesssim10^{-3}$ for our benchmark~\cite{Mingarelli_2018}, which can be safely neglected. This is the form of the GWB correlation typically used in PTA analyses~\cite{NANOGrav:2023gor}. 

By contrast, for a close pair with $d\lesssim d_\text{coh}(f)$, %
$\Delta \Phi_{ab}(f,\unit{\Omega})$ no longer varies rapidly with $\unit{\Omega}$, so the $PP$ term must be retained. Consider a pulsar pair far from the Earth, for which $d\ll d_\text{coh}(f)\ll L$. We expand the ORF to second order in $d/d_\text{coh}$ and drop geometric corrections of order $d/L$. The separation $\vec{d}$ then decomposes into longitudinal and transverse components,
\begin{equation}
\label{eqn:dparal}
    d_\parallel \equiv L_a-L_b, \qquad
    d_\perp \equiv (L_a+L_b)\sin\left(\zeta_{ab}/2\right),
\end{equation}
so that $d_\parallel$ is simultaneously the difference in the retarded times, \textit{i.e.} the $\unit{\Omega}$-independent part of $\Delta\Phi_{ab}$, and, up to $\mathcal{O}(\zeta_{ab}^2)$, the component of $\vec{d}$ along the line of sight to the midpoint between the pulsars. Aligning this line of sight with $\unit{z}$, so that $\unit{n}_{a,b}=(\pm\sin(\zeta_{ab}/2),\,0,\,\cos(\zeta_{ab}/2))$, 
the relative phase takes the form %
\begin{equation}
    \Delta \Phi_{ab} (f, \unit{\Omega}) \simeq 2\pi f\left[d_\parallel \left(1+\cos \theta \right) + d_\perp \sin \theta \cos \phi\right]/c. %
\end{equation}
Both antenna patterns can also be evaluated along $\unit{z}$, where the polarization sum collapses to
\begin{equation}
    \sum_{A} F_a^A(\unit{\Omega})\,F_b^A(\unit{\Omega})
    \simeq \frac{1}{4}\left(1-\cos\theta\right)^2 \, .%
\end{equation}
Here we work in the polarization basis $e^+_{ij}=m_i m_j - p_i p_j$ and $e^\times_{ij}=m_i p_j + p_i m_j$, where $\unit{m}$ and $\unit{p}$ are an orthonormal basis for the plane perpendicular to $\unit{\Omega}$, and we use the convention %
$\unit{\Omega}=(\sin \theta \cos \phi, \sin \theta \sin \phi, \cos \theta)$, $\unit{m}=(\sin \phi, -\cos \phi, 0)$ and $\unit{p}=(\cos \theta \cos \phi, \cos \theta \sin \phi, -\sin \theta)$~\cite{Romano:2023zhb}. %
Plugging this into Eq.~\eqref{eqn:GWB_ORF_short_wavelength}, and expanding to $\mathcal{O}((d/d_\text{coh}(f))^2)$, we find
\begin{equation} 
\begin{split}
    \Gamma^\text{GWB}_{ab} &\approx 1 +  \frac{3}{2}\int \frac{d^2\Omega}{4\pi }\frac{1}{4}(1-\cos \theta)^2  \left[e^{2\pi i f\left(d_\parallel(1+\cos \theta) + d_\perp \sin \theta \cos \phi\right)/c} -1\right] \\
    &\approx 1 +  \frac{3}{2}\int \frac{d^2\Omega}{4\pi }\frac{1}{4}(1-\cos \theta)^2  \left[\frac{2\pi i f d_\parallel(1+\cos \theta)}{c} + \frac{1}{2} \left[\frac{2\pi i f d_\parallel(1+\cos \theta)}{c}\right]^2 \right] \\
    & \qquad  +  \frac{3}{2}\int \frac{d^2\Omega}{4\pi }\frac{1}{4}(1-\cos \theta)^2  \times \frac{1}{2} \left[\frac{2\pi i f d_\perp\sin \theta \cos \phi}{c}\right]^2 \\
    &=1+\frac{i}{4} \left(\frac{d_\parallel}{d_\text{coh}(f)}\right)-\frac{1}{10} \left(\frac{d_\parallel}{d_\text{coh}(f)}\right)^2 -\frac{3}{40} \left(\frac{d_\perp}{d_\text{coh}(f)}\right)^2, %
\end{split}
\end{equation}
where terms with an odd power of $\cos\phi$ vanish upon the $\phi$ integration. We factorize the phase to rewrite this as
\begin{equation} \label{eqn:ORF_dparallel_neq_zero}
    \Gamma^\text{GWB}_{ab} \simeq e^{{i\pi f d_\parallel}/{(2c)}}\left[1-\frac{3}{40} \left(\frac{d_\text{eff}}{d_\text{coh}(f)}\right)^2 \right], \quad d_\text{eff}^2 = d_\perp^2 + \frac{11}{12}d_\parallel^2 \, ,
\end{equation}
which concludes our derivation of the ORF of the GWB in a form applicable to close pulsar pairs. In the main text, we assume all pulsars are at the same distance from the Earth, $L_a=L_b \equiv L$, and set $d_\parallel=0$ to obtain Eq.~\eqref{eqn:close_pulsar_ORF_schematic}. We now discuss the implications of relaxing this assumption.

\subsection{Dependence on pulsar pair orientation}
\label{supp:dependence_orientation}

\begin{figure*}
    \centering
    \includegraphics[width=\linewidth]{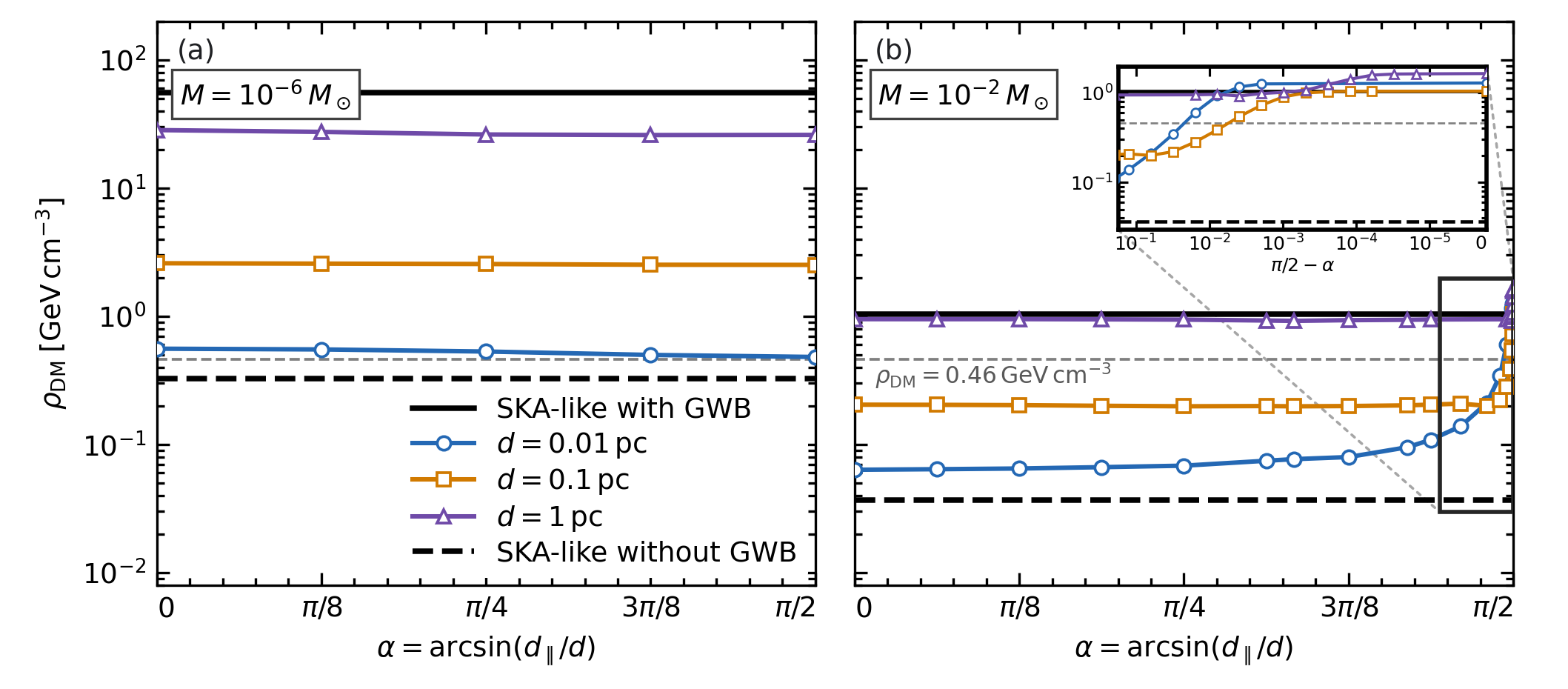}
    \caption{%
    The reach $\rho_\text{DM}$ as a function of the pair orientation angle, $\alpha\equiv \arcsin(d_\parallel/d)$, from transverse ($\alpha=0$) to longitudinal ($\alpha=\pi/2$), at fixed separations $d=0.01$, $0.1$, and $1\,\mathrm{pc}$ for the benchmark array in the main text. (a) At $M=10^{-6}M_\odot$, where the DM signal is dominated by the Doppler effect, the reach is essentially independent of the orientation. (b) At $M=10^{-2}M_\odot$, where the DM signal is instead dominated by the Shapiro delay, the reach is again essentially independent of orientation except for nearly longitudinal pairs, for which it degrades sharply.
}
    \label{fig:orientation_degradation}
\end{figure*}
Here we quantify how the reach depends on the pulsar pair's orientation relative to its line of sight, assuming that the pair separation $d$ is known. 
We parametrize this orientation by the angle $\alpha\equiv \arcsin(d_\parallel/d)$ and, in Fig.~\ref{fig:orientation_degradation}, plot the reach for our benchmark array of $100$ identical pulsar pairs as $\alpha$ varies from the transverse configuration ($\alpha=0$) assumed in the main text to a longitudinal configuration ($\alpha=\pi/2$). When $d_\parallel=0$, with our conventions, the ORF in Eq.~\eqref{eqn:ORF_dparallel_neq_zero} is real. However, rotating the pair away from $\alpha=0$ introduces a phase $e^{{i\pi f d_\parallel}/{(2c)}}$ into the ORF and the noise covariance from Eq.~\eqref{eqn:GWB_pair_covariance} changes to 
\begin{equation}
     \mathbf{s}(f)
    =
    \mathbf{I}_2
    +
    \left(\frac{f_\star}{f}\right)^\gamma
    \begin{pmatrix}
        1 & \left|\Gamma_{ab}(f)\right| e^{i\pi f d_\parallel/(2c)}  \\
       \left|\Gamma_{ab}(f)\right| e^{-i\pi f d_\parallel/(2c )} & 1
    \end{pmatrix}.
\end{equation}
The sum and difference channels are obtained by diagonalizing this covariance, and now become frequency dependent
\begin{equation}
    \widetilde{\delta t}_{\pm}(f)
    =
    \frac{1}{\sqrt{2}}
    \left[
        \widetilde{\delta t}_a(f)
        \pm
        e^{i\pi f d_\parallel/(2c)}
        \widetilde{\delta t}_b(f)
    \right].
\end{equation}
Notice that the sum and difference channel noise levels, $s_{\pm}(f) = 1+\left[1\pm |\Gamma_{ab}(f)|\right]\left({f_\star}/{f}\right)^\gamma$, remain unchanged, since they depend only on $|\Gamma_{ab}|$. %
The frequency dependence of the eigenvectors makes it difficult for us to compute the reach analytically, so we instead quantify its impact numerically in Fig.~\ref{fig:orientation_degradation} by plotting the reach to $\rho_{\mathrm{DM}}$ for a close pulsar-pair analysis as a function of $\alpha$, for some choices of $d$ and $M$. The numerics are described in Sec.~\ref{supp:numerics}. We find that at a Doppler-dominated benchmark mass $M=10^{-6}M_\odot$, the signal is local to each pulsar and the reach is essentially independent of orientation. The main effect is that rotating the pair from transverse to longitudinal shrinks $d_\text{eff}$ in Eq.~\eqref{eqn:ORF_dparallel_neq_zero} by a factor of $\sqrt{11/12}\simeq0.96$, marginally improving
the reach. At a Shapiro-dominated benchmark mass $M=10^{-2}M_\odot$, the reach again depends only weakly on orientation, unless the pair is nearly longitudinal. In this limit, the two pulsars lie along a common line of sight, so a subhalo imprints nearly identical Shapiro signals on both. This suppresses the difference-channel signal and drives the reach toward that of the sum channel. As this degradation occurs only near $\alpha =\pi/2$, an ensemble of pairs with random orientations is largely unaffected. We thus conclude that the simplification of setting $d_{\parallel}=0$ employed in the main text is well justified.

Note that we only observe the pulsars' angular positions. Together with an estimate of the distance to the system, one can determine the projected separation $d_\perp$, but not the line-of-sight separation $d_\parallel$. Consequently, observing $d_\perp<d_\text{coh}(f)$ does not by itself guarantee that the GWB is coherent across the pair, because $d_\parallel$, and hence $d_\text{eff}$, may still be large.
Marginalizing over an unconstrained $d_\parallel$ would likely cost significant sensitivity. However, in astrophysical environments like a globular cluster, additional information on  $d_\parallel$ can be obtained through signatures of the cluster's gravitational potential in the pulsars' timing data. Specifically, the line-of-sight acceleration and jerk induced by the cluster potential contribute to $\dot{\nu}$ and $\ddot{\nu}$, respectively, and have been used to infer pulsars' line-of-sight positions in Terzan~5~\cite{Prager:2016puh}. More speculatively, if the pair's GWB correlation were measurable, its magnitude and phase would encode $d_\text{eff}$ and $d_\parallel$, respectively, through Eq.~\eqref{eqn:ORF_dparallel_neq_zero}.

\section{Detecting a deterministic signal in a pulsar pair} %
\label{supp:sec:correlations_noise_eigenbasis}

In this section, we describe the expressions for the signal-to-noise ratio (SNR) used in the main text. %
Detecting a signal in a millisecond pulsar (MSP) requires modeling the various astrophysical phenomena that imprint on its timing data (see Ref.~\cite{Taylor:2021yjx} for a pedagogical introduction). In PTAs, these split into deterministic contributions, captured by the pulsar timing model, and stochastic contributions, which are treated as noise. One contribution to the timing model is the evolution of the pulsar's rotational phase, parametrized by its rotational frequency $\nu_a$ and spin-down rate $\dot{\nu}_a$, which enters the timing residuals as a quadratic polynomial in time. Pulsar timings are insensitive to any part of the signal that is degenerate with this timing model because the intrinsic parameters $\nu_a$ and $\dot{\nu}_a$ are not known independently and must be inferred from the data. There are other contributions to the timing model~\cite{Edwards:2006zg}, but the quadratic polynomial above is the dominant degeneracy for a DM signal and we restrict our attention to it.

For simplicity, we ignore the timing cadence and work in continuous time. We also make the simplifying assumption that each pulsar has the same total time of observation, which allows us to use the same timing model for each pulsar, with only the coefficients $\xi_a^{(i)}$ depending on the pulsar. For a given pulsar $a$ in the array, the timing residual can be written as %
\begin{equation}\label{eqn:delta_t_a}
    \delta t_a(t) = \sum_{i=0}^2 \xi_{a}^{(i)}  \phi_i (t)
    + \delta t_{\text{sig}, a}(t; \vec{\eta}) + n_a (t),
\end{equation}
where the $\phi_i$ %
form an orthonormal Legendre basis on the observing interval,
\begin{equation}\label{eqn:array_legendre_basis}
    \phi_i(t)
    =
    \sqrt{2n+1}\,
    P_i\!\left(\frac{2t}{T}-1\right),
    \qquad
    \frac{1}{T}\int_0^Tdt\,
    \phi_i(t)\phi_j(t)
    =
    \delta_{ij},
\end{equation}
with $P_i(x)$ being the $i$th Legendre polynomial. Here, $\delta t_{\text{sig}, a}$ is the signal of interest in pulsar $a$ with a deterministic template parametrized by $\vec{\eta}$, %
and $n_a(t)$ is the noise of the pulsar. One could also include a stochastic signal, such as the summed signal from many subhalos. However, in Ref.~\cite{Cherukupalli:2026cda} we found that for subhalos passing close to a pulsar or its line of sight, the stochastic signal reach in the Gaussian regime is subdominant to the deterministic signal search, and hence we omit it.

We assume that the noise is stationary and Gaussian, with a one-sided cross-PSD $S_{ab}(f)$, defined as
\begin{equation}\label{eqn:S_ab_def}
    \langle \widetilde{n}_a (f) \widetilde{n}^{\, *}_b(f')\rangle = \frac{1}{2} S_{ab}(f) \, \delta (f-f').
\end{equation}
In the pulsar basis, it is convenient to collect the residuals for all pulsars into a vector, denoted by $\delta\vec{t}(t)$, with components $\delta t_a(t)$. The expression of the cross-PSD matrix, denoted by $[\widehat{\mathbf{S}}(f)]_{ab}=S_{ab}(f)$ where the individual component is given by Eq.~\eqref{eqn:S_ab_def}, is shown in Eq.~\eqref{eqn:GWB_pair_covariance} for the pulsar pairs considered in the main text.

Our numerical results evaluate the SNR in this basis, working in the time domain as in Ref.~\cite{Cherukupalli:2026cda}, %
On a uniform cadence grid $t_i=i\Delta t$, with $i=0,\ldots,N_t-1$ and $N_t=\lfloor T/\Delta t\rfloor$, we collect the two residual time series into the $2N_t$-component vector
\begin{equation}
\delta\vec t_{{\rm sig},{\rm pair}}(\vec{\eta})
\equiv
\begin{pmatrix}
\delta\vec t_{{\rm sig},a}(\vec{\eta}) \\
\delta\vec t_{{\rm sig},b}(\vec{\eta})
\end{pmatrix}
\in\mathbb R^{2N_t}.
\label{eqn:DM_pair_signal_vector}
\end{equation}
We obtain the full $2N_t\times2N_t$ covariance $\mathbf{N}_{\rm pair}$ on the cadence grid by inverse Fourier transforming the pair cross-spectrum $S_{ab}(f)$ on the $N_t/2$ independent modes $f_k= k/T$. 

For each pulsar, we remove the component of its residuals that is degenerate with the quadratic timing model. Following Ref.~\cite{vanHaasteren:2012hj}, we collect the timing-model basis functions $\phi_i$ from Eq.~\eqref{eqn:array_legendre_basis} into the $N_t \times 3$ design matrix with monomial components $\mathbf{M}_i=[1, t_i, t_i^2]$. Its singular value decomposition provides $\mathbf G$, whose columns form an orthonormal basis for the subspace orthogonal to the timing model, \textit{i.e.}, ($\mathbf{G}^\top\mathbf{G}=\mathbf{I}_{N_t-3}$ %
and $\mathbf{G}^\top \mathbf{M} = 0$). Multiplication by $\mathbf{G}^\top$ is equivalent to projecting out the quadratic timing model. The two pulsars share the same sampling grid but have separate timing-model coefficients. For the stacked residual vector, the corresponding basis is therefore $\mathbf G_{\rm pair}\equiv \mathrm{diag}(\mathbf{G}, \mathbf{G})$. This construction of the projector would change if astrophysical effects lead to correlations in timing-model parameters between pulsars, as may be the case when the baryonic background in a globular cluster is jointly modeled. 

The covariance of the projected residuals is $\mathbf G_{\rm pair}^\top\mathbf N_{\rm pair}\mathbf G_{\rm pair}$. Its inverse defines the timing-marginalized inverse-noise operator \cite{vanHaasteren:2012hj,Cherukupalli:2026cda},
\begin{equation}
\mathbf N_{\perp,{\rm pair}}^{-1}
=
\mathbf G_{\rm pair}
\left(
\mathbf G_{\rm pair}^{\top}
\mathbf N_{\rm pair}
\mathbf G_{\rm pair}
\right)^{-1}
\mathbf G_{\rm pair}^{\top} \, .
\label{eqn:DM_pair_Nperp}
\end{equation}
The SNR then takes the form %
\begin{equation}
\operatorname{SNR}_{\rm pair}
^2(\vec{\eta})
=
\delta\vec t_{{\rm sig},{\rm pair}}^{\top}(\vec{\eta})\,
\mathbf N_{\perp,{\rm pair}}^{-1}\,
\delta\vec t_{{\rm sig},{\rm pair}}(\vec{\eta}) \, .
\label{eqn:DM_pair_snr}
\end{equation}

Our main results in Figs.~\ref{fig:sensitivity_gain} and \ref{fig:money_plot} use Eq.\eqref{eqn:DM_pair_snr} to compute the SNR numerically, applied to transiting DM substructure, as described in 
SM Sec.~\ref{supp:numerics}. However, to illustrate how the GWB contribution cancels in close pulsar pairs, it is convenient to work in a basis other than the pulsar basis. For simplicity, we consider the special case where $d_{\parallel}=0$. Let $\vec e_\alpha$ be eigenvectors of $\widehat{\mathbf{S}}(f)$, defined as
\begin{equation}\label{eqn:noise_eigenmodes}
    \widehat{\mathbf{S}}(f)\vec e_\alpha
    =
    S_\alpha(f)\vec e_\alpha,
    \quad
    \vec e_\alpha^\dagger\vec e_\beta
    =
    \delta_{\alpha\beta} \, ,
\end{equation}
which are independent of $f$ in this special case, as discussed in SM Sec.~\ref{supp:dependence_orientation}. Here $\alpha=1,2,\cdots,N_P$, where $N_P$ is the number of pulsars in the array.
We can then decompose the full timing residual in the noise eigenbasis as $\delta t_{\alpha}(t)=\vec e_\alpha^\dagger\delta\vec t(t)$. %
Because $\vec e_\alpha$ are independent of frequency, projecting onto them commutes with the Fourier transform, and the timing residual in each channel $\alpha$ is\footnote{We work with mean-subtracted residuals, which removes the constant mode $\phi_0$.}
\begin{equation}
    \delta t_\alpha(t) = \sum_{i=1}^2 \xi_{\alpha}^{(i)}  \phi_i (t)
    + \delta t_{\text{sig}, \alpha}(t) +n_\alpha (t),
\end{equation}
where $\xi_\alpha^{(i)} = \vec e_\alpha^\dagger\,\vec\xi^{(i)}$, $\delta t_{\text{sig},\alpha}(t) = \vec e_\alpha^\dagger\,\delta\vec t_{\text{sig}}(t)$, and the noise is uncorrelated across channels,
\begin{equation}
    \langle \widetilde{n}_\alpha (f) \widetilde{n}^{\, *}_\beta(f')\rangle = \frac{1}{2} S_\alpha (f) \, \delta_{\alpha\beta} \, \delta(f-f').
\end{equation}
These channels are then statistically independent: each channel behaves like a single pulsar with noise PSD $S_\alpha(f)$, to which the framework of Ref.~\cite{Cherukupalli:2026cda} can be directly applied. Following 
Ref.~\cite{Cherukupalli:2026cda}, we define the noise-weighted inner product in each channel by
\begin{equation}\label{eqn:noise_eigenmode_inner_product}
    (g|h)_\alpha
    \equiv
    4\,\mathrm{Re}\int_{1/T}^{\infty}df\,
    \frac{\widetilde g^{\,*}(f)\widetilde h(f)}{S_\alpha(f)}.
\end{equation}
Since the map $\xi_a^{(i)} \to \xi_\alpha^{(i)}$ is invertible, marginalizing over the timing model pulsar by pulsar is equivalent to marginalizing over it channel by channel. The SNR in each channel is then simply the norm of the projected signal, and the channel SNRs add in quadrature,
\begin{equation}
    \label{eqn:SNR_channel_decomp}
    \text{SNR}^2
    =
    \sum_\alpha\text{SNR}_\alpha^2,
    \quad
    \text{SNR}_\alpha^2
    \equiv
    \left(
        \delta t_{\mathrm{sig},\alpha}^{\perp}
        \middle|
        \delta t_{\mathrm{sig},\alpha}^{\perp}
    \right)_\alpha,
\end{equation}
where $\delta t_{\mathrm{sig},\alpha}^{\perp}$ denotes the signal after projecting out the timing model. Since stationarity of the noise and the opposite parity of $\phi_1$ and $\phi_2$ about the midpoint of the observing interval give $(\phi_1|\phi_2)_\alpha = 0$~\cite{Cherukupalli:2026cda}, this projection is 
\begin{equation}\label{eqn:appendix_projection_formula_channel}
    \delta t_{\mathrm{sig},\alpha}^{\perp}(t)
    =
    \delta t_{\mathrm{sig},\alpha}(t)
    -
    \sum_{i=1}^{2}
    \frac{(\phi_i|\delta t_{\mathrm{sig},\alpha})_\alpha}
    {(\phi_i|\phi_i)_\alpha}\phi_i(t).
\end{equation}

\section{Numerical computation of the reach}
\label{supp:numerics}

A subhalo worldline is parametrized by $\vec{\eta}\equiv(\vec r_0,\vec v)$, %
where $\vec{r}_0$ and $\vec{v}$ are its initial position and velocity, respectively. It imprints a residual in both pulsars of a pair through Eq.~\eqref{eqn:delta_t_t}. We stack these residuals together to form $\delta\vec t_\text{DM,pair}(\vec{\eta})$, %
as in Eq.~\eqref{eqn:DM_pair_signal_vector}, and evaluate the $\text{SNR}$ of the worldline through Eq.~\eqref{eqn:DM_pair_snr}. Note that in evaluating $S_{ab}(f)$, later transformed to $\mathbf{N}_\text{pair}$, we do not use the small-distance expansion of the ORF in Eq.~\eqref{eqn:ORF_dparallel_neq_zero}; rather, we numerically evaluate its full integral form in Eq.~\eqref{eqn:GWB_ORF_short_wavelength}.

We are interested in the number of events with an SNR above a given threshold $\text{SNR}_\text{th}$. This follows a Poisson distribution with mean
\begin{equation}
\bar{N}_\text{th}
=
\frac{\rho_{\rm DM} }{M}
\int d^3\vec r_0\,d^3\vec v\,
f_{\vec v}(\vec v)\,
\Theta\!\left[\operatorname{SNR}_{\rm pair}
(\vec{r}_0,\vec{v})-\text{SNR}_\text{th}\right],
\end{equation}
where $f_{\vec v}(\vec v)$ is the DM velocity distribution. For simplicity, we assume a fixed DM speed of $v=340\,\mathrm{km/s}$ with isotropically distributed velocity directions. The calculation can be generalized to a Maxwell--Boltzmann distribution~\cite{Ramani:2020hdo,Lee:2020wfn}.
To evaluate this integral, we follow Ref.~\cite{Cherukupalli:2026cda} and sample all subhalo worldlines that pass within a distance $R$ of either pulsar in a single pair. In Fig.~\ref{fig:money_plot}, we retain the Shapiro-dominated subhalos and also include all worldlines that pass within $R$ of either line of sight. Note again that we neglect subhalos passing close to the Earth because up to $\mathcal{O}(d/L)$ corrections, their signals are identical in both pulsars and vanish in the difference channel. 

Concretely, we first sample the velocity direction $\unit{v}$ isotropically, then %
sample the initial position $\vec{r}_0$ from the union of the Doppler and Shapiro capsules in Fig.~\ref{fig:dm_capsules}. We assign these capsules to have a radius $R$, calibrated independently for the Doppler and Shapiro cases. For a given $R$, we increase the number of samples until the quantities below converge. %
We choose $R$ large enough that increasing it further does not change the integral. The smallest $R$ for which the sampling region contains all worldlines with a nonzero contribution to the integrand $\Theta\!\left[\operatorname{SNR}(\vec{r}_0,\vec{v})-\text{SNR}_\text{th}\right]$ is the ``detection distance" defined in Ref.~\cite{Cherukupalli:2026cda}. 

In Fig.~\ref{fig:sensitivity_gain}, we compute the median $\text{SNR}$ of the loudest flyby near a \textit{single} pulsar pair, fixing $\rho_\text{DM}=0.46\, \mathrm{GeV/cm^3}$. That is, we only consider subhalos in the Doppler capsules. This can be rephrased as the largest $\text{SNR}_\text{th}$ such that in $50\%$ of realizations, at least one event exceeds it. In other words, since the event count follows a Poisson distribution, we solve $\bar{N}_\text{th}(\text{SNR}_\text{th}) = \ln 2$. We first sample many worldlines in these capsules, compute their SNRs, and report the threshold that satisfies this condition. 

In Fig.~\ref{fig:money_plot}, we compute the minimum $\rho_\text{DM}$ at which $90\%$ of realizations have at least one event that exceeds a fixed $\text{SNR}_{\rm th}=4$ threshold in an array of $100$ pulsar pairs. This can be rephrased as solving $\bar{N}_\text{th}(\rho_\text{DM})=\ln 10$ for the full array. We make the simplifying assumption that the pairs are separated far enough that any given flyby can at most affect one pair. We also drop any residual GWB correlation between distinct pairs that could otherwise be used to condition the noise. Under these assumptions, the pairs are effectively independent and we can simply rescale the reach of a single pulsar pair to $N_{P}/2$ pairs as
\begin{equation}
\rho_{\rm DM}%
=
\frac{2M\ln 10}{
\displaystyle N_P
\int d^3\vec{r}_0\,d^3\vec{v}\,
f_{\vec{v}}(\vec{v})\,
\Theta\!\left[ \operatorname{SNR}_{\rm pair}
(\vec{r}_0,\vec{v})-4
\right]} \, .
\end{equation}
Figure~\ref{fig:money_plot} of the main text and  Figs.~\ref{fig:orientation_degradation} and \ref{fig:timing_noise_degradation} show this reach. Fig.~\ref{fig:dop_static_reach} also shows this reach, only including subhalos in the Doppler capsules. %

. %

\section{Analytic estimate of the reach in the static Doppler limit}
\label{supp:dop_static}

\begin{figure}[!t]
    \centering
    \includegraphics[width=0.8\linewidth]{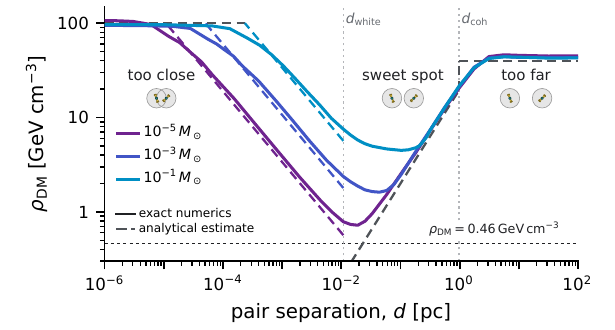}
    \caption{The minimum DM density in substructure $\rho_\text{DM}$ for which $90\%$ of random draws of the subhalo population contain at least one subhalo with $\mathrm{SNR}\geq 4$, as a function of the pair separation $d$ in an array of $100$ close pairs. We consider Doppler-dominated signals from subhalos sampled only within the pulsars' Doppler capsules. Solid curves show the numerical reach at subhalo masses $M=10^{-5},10^{-3},10^{-1}M_\odot$, and dashed curves show the corresponding analytical estimate in the static limit. Vertical guides mark the white-noise and GWB-coherence scales, $d_{\rm white}$ and $d_{\rm coh}(f=1/T)$. The reach improves linearly with decreasing pair separation in the \emph{sweet spot} relative to the \emph{too far} case. When the pulsars are \emph{too close}, their Doppler capsules, depicted as spheres, overlap significantly and the DM signal is tidally suppressed in the difference channel. The reach worsens as $d$ decreases until it plateaus at the sum-channel reach. %
    }
    \label{fig:dop_static_reach}
\end{figure}

The results in the main text, namely the sensitivity gain in Fig.~\ref{fig:sensitivity_gain} and the reach in $\rho_\text{DM}$ in Fig.~\ref{fig:money_plot}, are computed numerically as described in Sec.~\ref{supp:numerics}. In this section, we derive an analytical estimate of the reach
, which we summarize in Eq.~\eqref{eqn:doppler_static_reach_summary}. 
Figure~\ref{fig:dop_static_reach} compares this estimate with the numerical reach as a function of pair separation, for the benchmark array of $100$ pulsar pairs.
We consider only Doppler-dominated signals from subhalos passing close to the pulsars, and work in the static limit, $r_{\text{DM}} \gg vT$, where the subhalo moves little compared with its distance to the pulsar during the observation. We also take the two pulsars
at equal distances from Earth, $L_a=L_b\equiv L$, so that $d_\parallel$ defined in Eq.~\eqref{eqn:dparal} is zero. Under these assumptions, we evaluate the reach separately in the \emph{too far}, \emph{sweet spot}, and \emph{too close} regimes of pair seperation and find good agreement with the numerics wherever the assumptions hold.

Let the initial positions of the subhalo relative to these pulsars be $\vec{r}_{0, a}$ and $\vec{r}_{0, b}$, respectively. In the static limit, when $|\vec{r}_{0, a}|, |\vec{r}_{0, b}|\gg vT$, the subhalo-induced acceleration of each pulsar varies slowly over the observing window, and the
timing residual can be %
expanded in powers of $t$. The timing model is degenerate with terms up to $t^2$, so to leading order the signal can be expressed as a cubic polynomial, given by%
~\cite{Dror:2019twh, Cherukupalli:2026cda}
\begin{equation}
\label{eqn:static_doppler_signal_explicit}
\begin{split}
    \delta t_{a}( t)
    &\approx
    \frac{A}{r_{0, a}^3}
    \phi_3(t)
    \left[
        \unit{v}
        -3(\unit{r}_{0,a}\cdot\unit{v})\unit{r}_{0, a}
    \right]\cdot\unit{n}_a,\\
     \delta t_{b}( t)
    &\approx 
    \frac{A}{r_{0, b}^3}
    \phi_3(t)
    \left[
        \unit{v}
        -3(\unit{r}_{0,b}\cdot\unit{v})\unit{r}_{0, b}
    \right]\cdot\unit{n}_b,
\end{split}
\end{equation}
where $\phi_3$ is the cubic Legendre polynomial defined in Eq.~\eqref{eqn:array_legendre_basis} and $A\equiv GMvT^3/(120\sqrt{7} c)$. As noted in Eq.~\eqref{eqn:SNR_channel_decomp}, the SNR in each channel is set by the noise-weighted norm of the signal after projecting out the timing model,
\begin{equation} \label{eqn:snr_as_norm}
\begin{split}
    \mathrm{SNR}^2 &= \sum_{\alpha =\pm}\mathrm{SNR}^2_\alpha, \quad \mathrm{SNR}^2_\alpha = (\delta t_\alpha^\perp|\delta t_\alpha^\perp)_\alpha \equiv \frac{4}{S_\text{white}} \int_{1/T}^{\infty}\!df\,\frac{\big|\widetilde{\delta t}^\perp_{\alpha}(f)\big|^{2}}{s_{\alpha}(f)}, 
\end{split}
\end{equation}
where we assume that channel $\alpha$ has noise described by a one-sided PSD, $S_\alpha(f)\equiv S_\text{white}\, s_\alpha(f)$, with $S_\text{white} = 2\Delta t t^2_\text{rms}$. In the limiting regimes considered below, the noise in the sum and difference channels of a close pair takes the characteristic form 
\begin{equation} \label{eqn:spm_fstarpm_gamma_pm}
\begin{aligned}
    s_{\pm}(f)
    &= 1 + \left(\frac{f_\star}{f}\right)^\gamma
    \left[1 \pm |\Gamma_{ab}(f)|\right] \, ,\\
    &\simeq 1 + \left(\frac{f^{(\pm)}_\star}{f}\right)^{\gamma_{\pm}}.
\end{aligned}
\end{equation}
This allows us to use the results developed in Ref.~\cite{Cherukupalli:2026cda} to estimate the SNR analytically. The quantity of interest that enters Eq.~\eqref{eqn:snr_as_norm} is the inner product $(\phi_3^\perp|\phi_3^\perp)_\alpha$. Although the Legendre modes are orthonormal under the ordinary (white-noise-weighted) inner product in Eq.~\eqref{eqn:array_legendre_basis}, they are not under a red-noise-weighted inner product. In this case, the timing model mode $\phi_1$ and the static signal mode $\phi_3$ are not orthogonal. %
As a result, $\phi_3$ must itself be projected, and its norm in a channel with  $s_\alpha(f)=1+(f^{(\alpha)}_\star/f)^{\gamma_\alpha} $ takes two forms depending on whether the channel is dominated by white or red noise~\cite{Cherukupalli:2026cda},\footnote{Expressions for the noise-weighted norms of a DM signal in other  limits than the static Doppler limit can be found in Ref.~\cite{Cherukupalli:2026cda}.}
\begin{equation} \label{eqn:general_phi3_norm_scaling}
    (\phi_3^\perp|\phi_3^\perp)_\alpha
    \simeq \begin{cases}
    \displaystyle \frac{2T}{S_\text{white}}\,,&f^{(\alpha)}_\star T\ll1, \\[1em]
    \displaystyle\frac{6300}{\pi^6}
    \frac{T}{S_\text{white}} 
    \left(f^{(\alpha)}_\star T\right)^{-\min(\gamma_\alpha,5)},
    \quad
    &f^{(\alpha)}_\star T\gg1 \, .
    \end{cases}
\end{equation}
With Eq.~\eqref{eqn:static_doppler_signal_explicit} and Eq.~\eqref{eqn:general_phi3_norm_scaling} at hand, we now evaluate the reach for the static Doppel signal in each of the three separation regimes introduced above.
As in the main text, the reach is defined to be the smallest $\rho_{\rm DM}$ such that at least one subhalo has an $\mathrm{SNR}\geq4$ in $90\%$ of realizations.

\textbf{\emph{Too far}:} $d \gg \max(d_\text{coh}, r_\text{DM})$. For pulsar pairs with a large separation between the two pulsars, a subhalo can produce a significant signal in only one of the pulsars in the pulsar pair ($d\gg r_\mathrm{DM}$). The signals in the sum and difference channels are then $\delta t_- \approx \delta t_+ \approx \delta t_a/\sqrt{2}$, where $a$ is the pulsar that is perturbed by the subhalo. Since $d \gg d_{\text{coh}}$, the pulsar terms of the GWB  are uncorrelated and Eq.~\eqref{eqn:GWB_ORF_well_separated} gives $\Gamma_{ab}\approx \Gamma^\text{HD}_{ab}$. 
For 
$d\ll L$, the lines of sight are close to each other, $\unit{n}_a\approx \unit{n}_b \equiv \unit{n}$, such that  $\Gamma^\text{HD}_{ab}\approx \Gamma^\text{HD}_{aa} =1/2$. The noise parameters of $s_\pm(f)$ in Eq.~\eqref{eqn:spm_fstarpm_gamma_pm} become $f_\star^{(+)}=\left({2}/{3} \right)^{-{1}/{\gamma}}f_\star$ and  $f_\star^{(-)}=2^{-1/\gamma}f_\star$, with $\gamma_\pm=\gamma$. The SNR from Eq.~\eqref{eqn:snr_as_norm} thus takes the form
\begin{equation}
\label{eqn:SNRanalytic}
\begin{split}
    \mathrm{SNR}^2 &= \frac{A^2}{2r_{0, a}^6} \left[(\phi_3^\perp|\phi_3^\perp)_{-} + (\phi_3^\perp|\phi_3^\perp)_{+} \right]
    \,\Big|\left[
        \unit{v}
        -3(\unit{r}_{0,a}\cdot\unit{v})\unit{r}_{0, a}
    \right]\cdot\unit{n}_a\Big|^2.
\end{split}
\end{equation}
The angular factor in Eq.~\eqref{eqn:SNRanalytic} depends on the random directions of the subhalo's position and velocity. For a simple estimate, we replace it with its isotropic average,

\begin{equation}\label{eqn:static_doppler_angular_average}
    \left\langle\left|\left[\unit{v}-3(\unit{r}_{0, a}\cdot\unit{v})\unit{r}_{0, a}\right]\cdot\unit{n}_a\right|^2\right\rangle_{\unit r_{0, a},\unit v}=\frac{2}{3} \, ,
\end{equation}
so that
\begin{equation} \label{eqn:approx_SNR_local_flyby}
\begin{split}
    \mathrm{SNR}^2 &\simeq \frac{A^2}{3r_{0, a}^6} \left[(\phi_3^\perp|\phi_3^\perp)_{-} + (\phi_3^\perp|\phi_3^\perp)_{+} \right].
\end{split}
\end{equation}
Recall from Sec.~\ref{supp:sec:DM}
that 
$r_\text{DM}$ is the median distance of the closest flyby to any pulsar in the array. The reach, however, requires an SNR that is exceeded in $90\%$ of realizations. Since the SNR decreases monotonically with distance, we evaluate it at the closest-approach distance reached in $90\%$ of realizations, which is obtained by rescaling $r_\mathrm{DM}\to r_\mathrm{DM}\,(\ln 10/\ln 2)^{1/3}$. The expression for $r_{\text{DM}}$  in Eq.~\eqref{eqn:Doppler_rDM} assumes that the capsules of the two pulsars in a pair do not overlap, which holds here because $d \gg r_{\text{DM}}$.

For our noise benchmark parameters, $f_\star T \gg 1$, and as a result $f^{(+)}_\star T, \, f^{(-)}_\star T \gg 1$. We therefore apply the red-noise dominated branch in Eq.~\eqref{eqn:general_phi3_norm_scaling} for both channels, and evaluate the SNR at the rescaled $r_\mathrm{DM}$. Solving $\mathrm{SNR}(\rho_{\rm DM})\geq 4$ at the benchmark GWB parameters assumed in the main text, $\gamma=13/3$ and $A_\mathrm{GWB}=2.4\times10^{-15}$, we find

\begin{equation}\label{eqn:too_far_reach}
\boxed{\rho_\text{DM} \gtrsim 40\,\text{GeV/cm}^3 \left(\frac{200}{N_P}\right)\left(\frac{20\,\text{yr}}{T}\right)^{4/3}, \quad d \gg \max(d_\text{coh}, r_\text{DM}) \quad \text{(too far)}.} \
\end{equation}

This reach is independent of $M$ because the larger signal of a heavier subhalo is offset by the larger distance to the loudest flyby. It is also independent of $\Delta t$ and $t_\text{rms}$ because both channels are red-noise-dominated, so the GWB rather than the white noise sets the noise level.

\textbf{\emph{Sweet spot}:} $\max(r_{\rm DM},d_{\rm white})\ll d\ll d_{\rm coh}$. In this regime, the pair is close enough for the GWB to cancel in the difference channel, but not so close that the channel is white-noise dominated or the subhalo signal is affected. Since $d\gg r_\mathrm{DM}$, a subhalo can still only produce a detectable signal in one pulsar and the SNR is still given by Eq.~\eqref{eqn:approx_SNR_local_flyby}, similarly evaluated at the rescaled $r_\mathrm{DM}\to r_\mathrm{DM}\,(\ln 10/\ln 2)^{1/3}$.
However, the GWB is now suppressed in the difference channel, and the HD relation in Eq.~\eqref{eqn:GWB_ORF_well_separated} no longer applies. Instead, the leading-order expansion in the ORF is given by Eq.~\eqref{eqn:close_pulsar_ORF_schematic}, \textit{i.e.}, $\Gamma_{ab}(f)\approx 1 - \mathcal{O}\!\left[(d/d_\text{coh}(f))^2\right]$. Hence, in the difference channel, we have $\gamma_-=\gamma-2$ and $f_\star^{(-)}$ the crossover frequency in Eq.~\eqref{eqn:GWB_difference_crossover}, as highlighted in the main text. Since $d\gg d_\text{white}$, we have $f_\star^{(-)}T\gg1$, and the red-noise dominated branch of the SNR formula in Eq.~\eqref{eqn:general_phi3_norm_scaling} applies. The noise in the sum channel is much stronger, so $(\phi_3^\perp|\phi_3^\perp)_{+}  \ll  (\phi_3^\perp|\phi_3^\perp)_{-}$ and we can neglect the sum-channel contribution in Eq.~\eqref{eqn:approx_SNR_local_flyby}. For the same GWB parameters, we find that the reach is once again independent of the subhalo mass and the white noise parameters
\begin{equation}\label{eqn:sweet_spot_reach}
    \boxed{\rho_{\mathrm{DM}}
    \gtrsim 0.2\,\text{GeV}\,\text{cm}^{-3}
    \left(\frac{d}{0.01\, \mathrm{pc}}\right)\left(\frac{200}{N_P}\right)
    \left(\frac{20\,\text{yr}}{T}\right)^{7/3}, \quad \max(r_{\rm DM},d_{\rm white})\ll d\ll d_{\rm coh} \quad \text{(sweet spot)}.}
\end{equation}
The reach improves linearly as the pair separation decreases, as expected.

\textbf{\emph{Too close}:} $ d\ll\min(d_{\rm white}, r_\mathrm{DM}) $. In this regime, the pulsars are so close that the difference channel suppresses both the GWB and the DM signal and is dominated by white noise. Let $\vec{d}$ and $\vec{r}_0$ be the vector pointing from pulsar $a$ to pulsar $b$ and the initial position of the subhalo relative to the midpoint of the pair, respectively. Then $\vec r_{0,a}=\vec r_0+\vec d/2$ and $\vec r_{0,b}=\vec r_0-\vec d/2$. Once again, up to $\mathcal O(d/L)$ corrections, we set $\unit n_a\simeq\unit n_b\equiv\unit n$. For a tidal encounter, $d\ll r_0$, and we can expand the signal in Eq.~\eqref{eqn:static_doppler_signal_explicit} about the pair midpoint and form the sum and difference channels. We find
\begin{subequations}\label{eqn:tidal_channel_expansion}
\begin{align}
    \delta t_{+}(t)
    &\simeq\frac{\sqrt{2} A}{r_{0}^3}
    \phi_3(t)
    \left[
        \unit{v}
        -3(\unit{r}_{0}\cdot\unit{v})\unit{r}_{0}
    \right]\cdot\unit{n},
    \\
    \delta t_{-}(t)
    &\simeq
    \frac{A}{\sqrt{2}} \phi_3(t)
    (\vec d\cdot\nabla_{\vec{r}_0})\left[\frac{1}{r_0^3} \left(
            \unit v-3(\unit{r}_0\cdot\unit{v})\unit{r}_0
        \right)\cdot\unit{n}\right].
\end{align}
\end{subequations}
The difference-channel signal is tidally suppressed by a factor of order $d/r_0$. As a result, at small enough separation, the sum channel is eventually the main contributor to the SNR. After averaging over $\unit{r}_0$ and $\unit{v}$ assuming isotropic distributions, the SNR in the sum channel takes a form similar to that in Eq.~\eqref{eqn:approx_SNR_local_flyby}, 
\begin{equation}
    \mathrm{SNR}_+^2 \simeq \frac{4A^2}{3r_{0}^6}  (\phi_3^\perp|\phi_3^\perp)_{+}.
\end{equation}
The noise parameters in the sum channel are given by$f_\star^{(+)}=2^{1/\gamma}f_\star$ and $\gamma_+=\gamma$. For our noise benchmark, $f_\star^{(+)}T \gg 1$, so again we use the red noise dominated branch to compute the SNR from
Eq.~\eqref{eqn:general_phi3_norm_scaling}. On the other hand, in the difference channel, the SNR takes the form
\begin{equation} \label{eqn:SNR_minus_tidal_expanded}
    \mathrm{SNR}^2_- \simeq \frac{A^2}{2} (\phi_3^\perp|\phi_3^\perp)_{-} \, \left|
            (\vec d\cdot\nabla_{\vec r_0})
            \left[
                \frac{1}{r_0^3}
                \left(
                    \unit v-3(\unit r_0\cdot\unit v)\unit r_0
                \right)\cdot\unit n
            \right]
        \right|^2.
\end{equation}
In this regime $d\ll d_\text{white}$, such that  $f_\star^{(-)}T \ll 1$ and we use the white-noise dominated branch of Eq.~\eqref{eqn:general_phi3_norm_scaling} to compute the SNR. 

We once again replace the angular factor with its isotropic average to simplify Eq.~\eqref{eqn:SNR_minus_tidal_expanded}, which we calculate as follows. First, we perform the directional derivative on Eq.~\eqref{eqn:tidal_channel_expansion}, finding
\begin{equation}\label{eqn:tidal_signal_derivative}
\begin{split}
    &(\vec d\cdot\nabla_{\vec{r}_0})
    \left[
        \frac{1}{r_0^3}
        \left(
            \unit v-3(\unit{r}_0\cdot\unit{v})\unit{r}_0
        \right)\cdot\unit{n}
    \right]
    \\
    &\qquad=
    \frac{3d}{r_0^4}
    \left\{
        \left[
            5(\unit{r}_0\cdot\unit{d})(\unit{r}_0\cdot\unit{v})
            -(\unit{d}\cdot\unit{v})
        \right]\unit{r}_0
        -(\unit{r}_0\cdot\unit{d})\unit{v}
        -(\unit{r}_0\cdot\unit{v})\unit{d}
    \right\}\cdot\unit{n} \, ,
\end{split}
\end{equation}
which allows us to compute the angular average by directly integrating over $\unit r_0$ and $\unit v$ assuming isotropic distributions. This gives
\begin{equation}\label{eqn:tidal_derivative_angular_average}
    \left\langle
        \left|
            (\vec d\cdot\nabla_{\vec{r}_0})
            \left[
                \frac{1}{r_0^3}
                \left(
                    \unit v-3(\unit{r}_0\cdot\unit{v})\unit{r}_0
                \right)\cdot\unit{n}
            \right]
        \right|^2
    \right\rangle_{\unit r_0,\unit v}
    =
    \frac{d^2}{r_0^8}
    \left[
        3+(\unit d\cdot\unit n)^2
    \right].
\end{equation}
Since we assume $d_\parallel=0$, which implies $\unit d\cdot\unit n\approx 0$, we substitute Eq.~\eqref{eqn:tidal_derivative_angular_average} into Eq.~\eqref{eqn:SNR_minus_tidal_expanded}, and find
\begin{equation} 
    \mathrm{SNR}^2_- \simeq \frac{3A^2d^2}{2r_0^8} (\phi_3^\perp|\phi_3^\perp)_{-}.
\end{equation}
Note, however, that in this limit $d \ll r_{\rm DM}$, and the two Doppler capsules of the pair almost fully overlap. As discussed in Sec.~\ref{supp:sec:DM}, this means that the total volume now instead scales with the number of pairs, $N_P/2$, and the rescaled $90\%$ distance is instead
$r_\mathrm{DM}\to r_\mathrm{DM}\,(2\ln 10/\ln 2)^{1/3}$. Evaluating each $\mathrm{SNR}$ at the rescaled $r_\mathrm{DM}$ and solving $\mathrm{SNR}_{\pm}( \rho^{(\pm)}_{\mathrm{DM}})\geq4$ in each channel separately, we obtain
\begin{empheq}[box=\fbox]{align}    \rho^{(-)}_{\mathrm{DM}}&\gtrsim 0.6\,\text{GeV}\,\text{cm}^{-3}
    \left(\frac{M}{10^{-5}M_\odot}\right)^{1/4}
    \left(\frac{0.01\,\rm{pc}}{d}\right)^{3/4} \left(\frac{200}{N_P}\right)
\left(\frac{20\,\mathrm{yr}}{T}\right)^{21/8}
\left(\frac{\Delta t}{2\,\mathrm{wk}}\right)^{3/8}
\left(\frac{t_{\rm rms}}{50\,\mathrm{ns}}\right)^{3/4},
    \label{eqn:too_close_reach_-}\\
    \rho^{(+)}_{\mathrm{DM}}
    &\gtrsim
    100\,\text{GeV}\,\text{cm}^{-3} \left(\frac{200}{N_P}\right)
    \left(\frac{20\,\text{yr}}{T}\right)^{4/3}, \quad d\ll\min(d_{\rm white}, r_\mathrm{DM})\quad \text{(too close).}
    \label{eqn:too_close_reach_+}
\end{empheq}

Note that here the difference channel has a reach that now depends on both the subhalo mass and white-noise parameters. Its reach worsens with decreasing pair separation, as the DM signal continues to be tidally suppressed against a $d$-independent white noise floor. As a simple analytical envelope, we take the reach to be set by the more sensitive channel, $ \rho_{\mathrm{DM}} \gtrsim \min( \rho^{(-)}_{\mathrm{DM}},\,\rho^{(+)}_{\mathrm{DM}})$. We define $d_\text{sum}(M)$ as the separation at which the reach crosses over into the sum channel, $\rho^{(+)}_{\mathrm{DM}}=\rho^{(-)}_{\mathrm{DM}}$, which in Fig.~\ref{fig:dop_static_reach} is marked by the transition from the colored dashed lines with a linear slope to a flat line, and depends on $M$.

Finally, putting Eqs.~\eqref{eqn:too_far_reach}, \eqref{eqn:sweet_spot_reach}, and \eqref{eqn:too_close_reach_-}--\eqref{eqn:too_close_reach_+} together, with all other parameters fixed to their benchmark values, we find
\begin{equation}
\label{eqn:doppler_static_reach_summary}
\boxed{\begin{gathered}
\text{\normalfont\bfseries Analytic estimate: Doppler signal in the static limit}\\
\rho_{\rm{DM}} (M, d) \gtrsim
\begin{cases}
\displaystyle
40\,\mathrm{GeV\,cm^{-3}},
&
d\gg \max(d_{\rm coh}, r_{\rm DM}),
\\[6pt]
\displaystyle
0.2\,\mathrm{GeV\,cm^{-3}}
\left(\frac{d}{0.01\, \mathrm{pc}}\right),
&
\max(r_{\rm DM},d_{\rm white})\ll d\ll d_{\rm coh},
\\[8pt]
\displaystyle
0.6\,\mathrm{GeV\,cm^{-3}}
\left(\frac{M}{10^{-5}M_\odot}\right)^{1/4}
\left(\frac{0.01\, \mathrm{pc}}{d}\right)^{3/4},
&
d_{\rm sum}\ll d\ll\min(d_{\rm white},r_{\rm DM}),
\\[8pt]
\displaystyle
100\,\mathrm{GeV\,cm^{-3}},
&
d\ll d_{\rm sum}.
\end{cases}
\end{gathered}}
\end{equation}
This is the analytical estimate overlaid as dashed lines on the numerical results shown in solid lines in Fig.~\ref{fig:dop_static_reach}. Since $r_\text{DM}$ depends on $\rho_\text{DM}$, we use the numerical reach curve $\rho_\text{DM}(M, d)$ to compute $r_\text{DM}(M, d)$, apply the rescalings above, and check the conditions defining each regime, along with the validity of the static limit. %

The estimate reproduces the numerical reach within each regime and deviates mainly near the transitions between them. The reach is best near the lower edge of the sweet spot, $d \sim \max(r_\text{DM}, d_\text{white})$, which for our benchmark falls at $d \sim 0.01$--$0.1\,$pc depending on $M$. For $M = 10^{-5}\,M_\odot$, the static limit breaks down.

\section{The difference channel as a high-pass filter of angular modes}
\label{supp:continuum_angular_structure}

The goal of this section is to have a description of the GWB as well as the DM substructure in terms of angular modes; we will see that the characteristic angular scale of dark matter substructure is very different from the GWB, giving a handle to separate the two.  

To illustrate the point, we work with an idealized sphere of pulsars around the Earth with radius $L=5\, \rm{kpc}$. We evaluate the response on it from the GWB, as well as the Doppler signal from a characteristic subhalo flyby at a distance $r_\mathrm{DM}\gg vT$ from the sphere. We then decompose these responses into angular modes, given in Eq.~\eqref{eqn:GWB_angular_spectrum_regimes} for the GWB and Eq.~\eqref{eqn:static_doppler_angular_spectrum_approx} for the DM signal, to develop an $\ell$-space picture of close-pair filtering that complements Fig.~\ref{fig:sensitivity_gain}. 

\begin{figure}[!t]
    \centering
    \includegraphics[width=\linewidth]{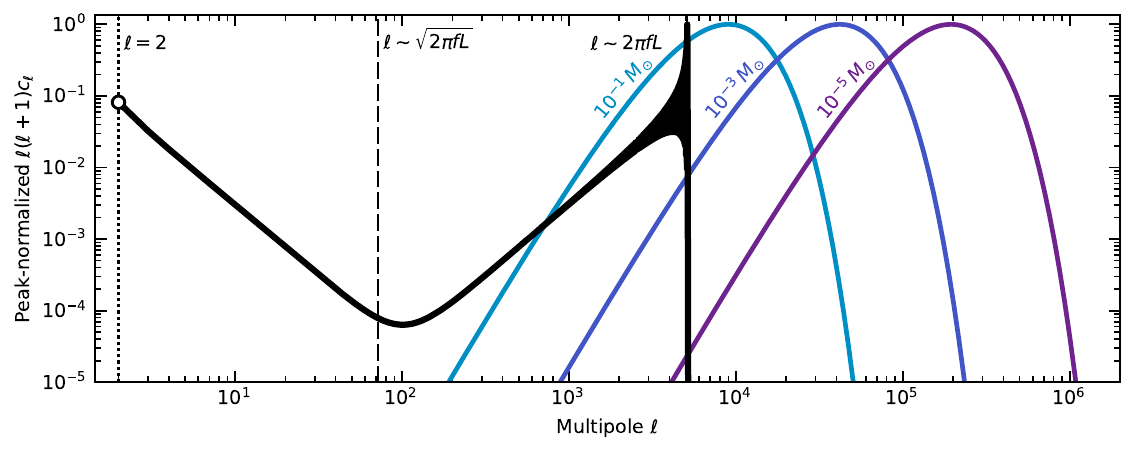}
    \caption{Angular power $\ell(\ell+1)c_\ell$ of the GWB (black), evaluated at $f=1/T$, and a characteristic Doppler event with the indicated subhalo masses (colored). Each curve is normalized to its peak independently, so the figure compares only their angular structure. A close pair leverages the fact that a DM signal's angular power can extend much further than the GWB's cutoff at $\ell\sim 2\pi fL$. See text for additional discussion.
    }
    \label{fig:continuum_spectra_DM_GWB}
\end{figure}

We will describe the procedure in more detail below, but first summarize the result in Fig.~\ref{fig:continuum_spectra_DM_GWB}, where we plot their angular power $\ell(\ell+1)c_\ell$ as a function of $\ell$ and see that the GWB spectrum becomes suppressed near $\ell_{\rm GWB}\sim2\pi fL$, whereas the DM signal extends to much larger multipoles $\ell_\text{DM}\sim L/r_\text{DM}$. A close pair exploits this difference in angular structure without reconstructing the individual angular modes in Fig.~\ref{fig:continuum_spectra_DM_GWB}.  
Instead, a close pair separated by distance $d$ makes a two-point measurement at $\unit{n}_a$ and $\unit{n}_b$ across $\Delta\theta\simeq d/L$. The difference channel of this measurement then acts as a high-pass filter on angular structure: it suppresses variations on scales larger than its separation, and allows signals with structure on scales smaller than its separation to pass. To see this, notice that the contribution of each angular mode $(\ell,m)$ to the difference channel is weighted by $Y_{\ell m}(\unit{n}_a)-Y_{\ell m}(\unit{n}_b)$. Averaging its square magnitude over $m$, we find 
\begin{equation}
\begin{split}
    \frac{1}{2\ell+1} \sum_{m=-\ell}^{\ell} |Y_{\ell m}(\unit{n}_a) -Y_{\ell m}(\unit{n}_b)|^2 &= \frac{1}{2\pi} \left[1-P_\ell(\cos\Delta\theta)\right]. %
\end{split}
\end{equation}
In the limit $\ell\Delta \theta\ll 1$, we have $1-P_\ell(\cos\Delta\theta) \approx (\ell \Delta\theta)^2/4$. Therefore, angular modes with $\ell\ll 1/\Delta \theta$ are quadratically suppressed.  When $d\ll d_{\rm coh}(f)$, even the finest GWB structure satisfies $\ell_{\rm GWB}\Delta\theta\sim d/d_{\rm coh}(f)\ll1$ and is suppressed. By contrast, the DM signal can extend to much higher multipoles $\ell_\text{DM}\Delta \theta \gtrsim 1$ and survives, as we now derive.

\subsection{Angular structure of the GWB}
\label{supp:GWB_spectra}

\label{supp:close_pair_filtering}

Considering the response of isotropically distributed pulsars on a sphere of radius $L$ to an isotropic GWB, the ORF admits a rotationally invariant expansion
\begin{equation}\label{eqn:GWB_angular_spectrum}
    \Gamma^\text{GWB}_{ab}(f)
    =
    \sum_{\ell=0}^{\infty}\sum_{m=-\ell}^{\ell}
    c_\ell^{\rm GWB}(f)\,
    Y_{\ell m}(\unit n_a)Y_{\ell m}^*(\unit n_b).
\end{equation}
Projecting onto this basis, substituting Eq.~\eqref{eqn:GWB_ORF_exact}, exchanging the order of integration, and averaging over the $2\ell+1$ equivalent values of $m$, we obtain:%
\begin{equation}\label{eqn:GWB_angular_spectrum_response}
\begin{split}
    c_\ell^{\rm GWB}(f)
    &=\frac{1}{2\ell + 1}\sum_{m=-\ell}^{\ell}
    \int d^2\unit{n}_a\,d^2\unit{n}_b\,
    Y_{\ell m}^*(\unit{n}_a)\,
    \Gamma^\text{GWB}_{ab}(f)\,
    Y_{\ell m}(\unit{n}_b)\\
    &=
    \frac{3}{2(2\ell + 1)}
    \int\frac{d^2\unit{\Omega}}{4\pi}
    \sum_{A=+,\times}\, \sum_{m=-\ell}^{\ell}\, \left|
    \int d^2\unit{n}\,
    F^A(\unit{n},\unit{\Omega})
    \left[
        1-e^{2\pi i fL(1+\unit{\Omega}\cdot\unit{n})}
    \right]
    Y_{\ell m}(\unit{n})
    \right|^2 .
\end{split}
\end{equation}
Here $F^A(\unit{n},\unit{\Omega})$ is the continuum version of the antenna pattern defined in Eq.~\eqref{eqn:antenna_pattern}:
\begin{equation}
    F^A(\unit{n},\unit{\Omega}) = \frac{\unit{n}^i \unit{n}^j }{2(1+\unit{\Omega}\cdot \unit{n})}e_{ij}^A (\unit{\Omega}).
\end{equation}
After summing over $m$ and $A$, the integrand of the $\unit{\Omega}$ average in Eq.~\eqref{eqn:GWB_angular_spectrum_response} is rotationally invariant and hence independent of $\unit{\Omega}$. We choose $\unit{\Omega}=\unit{z}$ and evaluate the angular integral
\begin{equation}\label{eqn:c_ell_zhat_expr}
\begin{split}
    c_\ell^{\rm GWB}(f)
    &=
    \frac{3}{2(2\ell + 1)}
    \sum_{A=+,\times}\, \sum_{m=-\ell}^{\ell}\, \left|
    \int d^2\unit{n}\,
    F^A(\unit{n},\unit{z})
    \left[
        1-e^{2\pi i fL(1+\unit{z}\cdot\unit{n})}
    \right]
    Y_{\ell m}(\unit{n})
    \right|^2 \, ,
\end{split}
\end{equation}
where we use the following spherical-harmonic convention
\begin{equation}\label{eqn:spherical_harmonic_convention}
    Y_{\ell m}(\theta,\phi)
=
\sqrt{\frac{2\ell+1}{4\pi}\frac{(\ell-m)!}{(\ell+m)!}}\,
P_\ell^m(\cos\theta)e^{im\phi},
\end{equation}
and write $\unit{n}=(\theta,\phi)$ in this frame. The inner integral then becomes
\begin{equation} \label{eqn:zhat_expr}
\begin{split}
    &\int d^2\unit{n}\,
    F^A(\unit{n},\unit{z})
    \left[
        1-e^{2\pi i fL(1+\unit{z}\cdot\unit{n})}
    \right]
    Y_{\ell m}(\unit n)
    \\
    &=
        \sqrt{
        \frac{2\ell+1}{4\pi}
        \frac{(\ell-m)!}{(\ell+m)!}
    }
    \int_{-1}^1 d\cos\theta\,
    \left[
        1-e^{2\pi i fL(1+\cos\theta)}
    \right]
    P_\ell^{m}(\cos\theta)
    \int_0^{2\pi} d\phi\,
    F^A(\theta,\phi)e^{im\phi}.
\end{split}
\end{equation}
In addition, we choose the polarization tensors $e^+_{ij}=x_i x_j-y_i y_j$ and $e^\times_{ij}=x_i y_j+y_i x_j$. The corresponding antenna patterns are $F^+(\theta,\phi)=\frac{1}{2}(1-\cos\theta)\cos2\phi$ and $F^\times(\theta,\phi)=\frac{1}{2}(1-\cos\theta)\sin2\phi$, whose azimuthal integrals are
\begin{equation} \label{eqn:azimuthal_integrals}
\begin{split}
    \int_0^{2\pi}d\phi\,
    F^+(\theta,\phi)e^{im\phi}
    &=
    \frac{\pi}{2}(1-\cos\theta)
    \left(\delta_{m,2}+\delta_{m,-2}\right),\\
    \int_0^{2\pi}d\phi\,
    F^\times(\theta,\phi)e^{im\phi}
    &=
    \frac{i\pi}{2}(1-\cos\theta)
    \left(\delta_{m,2}-\delta_{m,-2}\right).
\end{split}
\end{equation}
Since these $m=\pm2$ modes exist only for $\ell\geq2$, we find $c_0^{\rm GWB}=c_1^{\rm GWB}=0$, reflecting the spin-$2$ nature of GWs. %
Substituting Eq.~\eqref{eqn:azimuthal_integrals} into Eq.~\eqref{eqn:zhat_expr} and then summing over $m$ and $A$ in Eq.~\eqref{eqn:c_ell_zhat_expr} gives
\begin{equation}\label{eqn:GWB_angular_spectrum_exact}
    c_\ell^{\rm GWB}(f)
    =
    c_\ell^{\rm HD}
    \left|q_\ell(2\pi fL)\right|^2,
\end{equation}
where $c_\ell^{\rm HD}$ is the Hellings--Downs angular spectrum,
\begin{equation}\label{eqn:HD_correlation}
    c_\ell^{\rm HD}=6\pi\frac{(\ell-2)!}{(\ell+2)!}=\frac{6\pi}{(\ell+2)(\ell+1)\ell(\ell-1)}, \, \ell\geq2 \, ,
\end{equation}
and the finite-distance response factor, $q_\ell(2\pi fL)$, is defined by
\begin{equation}\label{eqn:GWB_qell_definition}
    q_\ell(x)
    \equiv
    \frac{1}{4}
    \int_{-1}^{1}d\mu\,
    (1-\mu)
    \left[
        1-e^{ix(1+\mu)}
    \right]
    P_\ell^{m=2}(\mu) \,.
\end{equation}
To evaluate this integral, we substitute the identity $P_\ell^{m=2}(\mu)=2\mu P_\ell'(\mu)-\ell(\ell+1)P_\ell(\mu)$
into Eq.~\eqref{eqn:GWB_qell_definition} and integrate the term containing $P_\ell'$ by parts. For $\ell \geq 2$, orthogonality of Legendre polynomials gives $\int_{-1}^{1}d\mu \, P_\ell(\mu)=\int_{-1}^{1}d\mu \, \mu P_\ell(\mu) = 0$. Defining $\Lambda_\ell=\ell(\ell+1)$, we then find
\begin{equation}
    q_\ell(x)=\frac{e^{ix}}{4} \int_{-1}^{1} d\mu \left[\Lambda_\ell+2 -\mu(\Lambda_\ell+4)+2ix \mu(1-\mu)\right] e^{ix\mu} P_\ell(\mu).
\end{equation}
Using the identity $\int_{-1}^{1} d\mu \, e^{ix\mu} P_\ell(\mu)=2i^\ell j_\ell(x)$ and its derivative with respect to $x$ to evaluate higher moments, the integral simplifies to
\begin{equation}\label{eqn:GWB_qell_bessel}
\begin{split}
    q_\ell(x)
    =
    \frac{1}{2}i^\ell e^{ix}\left[g_\ell(x)+ig'_\ell(x)\right], \quad g_\ell(x)=(\Lambda_\ell+2) j_\ell(x)+2x j'_\ell(x).
\end{split}
\end{equation}
In the limit where the GW wavelength is shorter than the Earth-pulsar distance, %
$x \equiv 2\pi fL \gg1$, there are three regimes of $\ell$ in which the spectrum simplifies:
\begin{enumerate}[label=\arabic*.]
    \item In the regime $2\leq\ell\ll\sqrt{x}$, we have $\Lambda_\ell \ll x$ and $g_\ell(x)\approx 2xj_\ell'(x) + \mathcal{O}(\ell^2/x)$. Using the large $x$ asymptotic $j_\ell'(x)\sim \cos(x-\pi \ell/2)/x$, we find $g_\ell(x)\approx 2\cos (x-\pi \ell/2)$ and subsequently, $|q_\ell|^2\approx 1$.   
    \item For $\sqrt{x}\ll \ell\ll {x}$, we have $\Lambda_\ell \gg x$ and $g_\ell(x)\approx \ell^2 j_\ell(x) + \mathcal{O}(x/\ell^2)$. Using the large $x$ asymptotic $j_\ell(x)\sim \sin(x-\pi \ell/2)/x$, we find $g_\ell(x)\approx (\ell^2/x) \, \sin (x-\pi \ell/2)$ and subsequently, $|q_\ell|^2\approx \ell^4/(4x^2)$.
    \item In the limit $\ell \gg x$, we have the asymptotic
    \begin{equation}
        j_\ell(x) \sim \frac{1}{2\sqrt{2}\ell}\left(\frac{ex}{2\ell}\right)^\ell .
    \end{equation}
    In this regime we still have $g_\ell(x) \approx \ell^2 j_\ell(x)$. $g'_\ell(x)$ dominates $g_\ell(x)$ parametrically in $\ell/x$ and we find
    \begin{equation}
        |q_\ell(x)|^2 \approx \frac{1}{4} \left[g'_\ell(x)\right]^2 \sim \frac{\ell^4}{32x^2}\left(\frac{ex}{2\ell}\right)^{2\ell}.
    \end{equation}
\end{enumerate}
Substituting these three asymptotic forms of $|q_\ell(x)|^2$ into Eq.~\eqref{eqn:GWB_angular_spectrum_exact}, and using $c_\ell^\text{HD}\approx6\pi/\ell^4$ in the latter two regimes, gives
\begin{equation}\label{eqn:GWB_angular_spectrum_regimes}
    \boxed{c_\ell^{\rm GWB}(f)
    \simeq
    \begin{cases}
        \displaystyle
        \frac{6\pi}{(\ell+2)(\ell+1)\ell(\ell-1)},
        & 2\leq\ell\ll\sqrt{2\pi fL},
        \\[10pt]
        \displaystyle
        \frac{3\pi}{2(2\pi fL)^2},
        & \sqrt{2\pi fL}\ll\ell\ll2\pi fL,
        \\[10pt]
        \displaystyle
        \frac{3\pi}{16(2\pi fL)^2} \left(\frac{e\pi fL}{\ell}\right)^{2\ell},
        & \ell\gg2\pi fL \, .
    \end{cases}}
\end{equation}
We see that the GWB's angular spectrum follows the Hellings--Downs spectrum at low multipoles, transitions near $\ell\sim\sqrt{2\pi fL}$ to a constant plateau $c_\ell^{\rm GWB}\sim 1/(2\pi fL)^2$, and is exponentially suppressed beyond $\ell\sim2\pi fL$. Note that this finite-distance response has been obtained previously in several complementary formalisms~\cite{Gair_2015,Ng_2022,Bernardo_2023,andrianov2026gravitationalwaveskymappingpulsar}.

As an additional comment, one can observe that the GWB's continuum power from Eq.~\eqref{eqn:GWB_angular_spectrum_exact} starts at $\ell=2$, and might naively conclude that dipolar signals ($\ell=1$), such as those that arise from subhalos passing close to Earth, occupy a GWB-free channel. However, in any finite array, the spherical harmonics are not perfectly orthogonal and can thus leak into each other. Working in the pulsar basis, one would observe that including these subhalos does not meaningfully change the reach and the GWB would still greatly limit DM sensitivity in a conventional array~\cite{Foster:2026kfg}.

\subsection{Angular structure of a DM flyby}

Having derived the spherical harmonic expansion of the GWB for close pulsar pairs in Eq.~\eqref{eqn:GWB_angular_spectrum_regimes}, we now derive the corresponding expansion for the DM signal. We consider the Doppler-dominated signal from a characteristic subhalo flyby that is initially at a distance $r_\text{DM}$, defined in Eq.~\eqref{eqn:Doppler_rDM} with $N_P=1$, from the pulsar sphere. %
As in Sec.~\ref{supp:dop_static}, we work in the static limit $r_\text{DM}\gg vT$, where projecting out the timing model leaves a cubic signal (Eq.~\eqref{eqn:static_doppler_signal_explicit}) that is set by the initial time derivative of the pulsar's acceleration along its line of sight. We call this the line-of-sight jerk, $j_\parallel$. We evaluate $j_\parallel(\unit{n})$ on the pulsar sphere and decompose it into its angular multipoles  $j_{\parallel, \ell m}$ . Unlike an isotropic GWB, a single DM event is manifestly anisotropic. Its signal is strongest at the point on the pulsar sphere closest to the subhalo, and the direction of its velocity $\unit{v}$ introduces an additional anisotropy. Nevertheless, we define a schematic angular spectrum $c_\ell^\text{DM}$ by averaging $|j_{\parallel, \ell m}|^2$ over $\unit{v}$ and the $2\ell + 1$ values of $m$.

We first define the coordinate system. Let $\vec{R}\equiv \vec{R}(0) = R \, \unit{R}$ be the initial position of the subhalo relative to the Earth, so that $\vec{R}(t)= \vec{R}(0)+\vec{v}t$ is its trajectory. We also define $\vec r(\unit{n},t)\equiv\vec R(t)-L\unit{n}$ as the position of the subhalo relative to a pulsar located at $\unit{n}$ on the sphere. Note that as the subhalo is at a distance $r_\text{DM}$ from the sphere at $t=0$, we have $|R-L|=r_\mathrm{DM}$.

The subhalo-induced acceleration of the pulsar along its line of sight $\unit{n}$ can be conveniently written as
\begin{equation}
    a_\parallel(\unit n,t)
    =
    GM\frac{\vec r(\unit n,t)\cdot\unit n}
    {|\vec r(\unit n,t)|^3}
    =
    GM\,\partial_L
    \left[
        \frac{1}{|\vec R(t)-L\unit n|}
    \right],
\end{equation}
where $\partial_L$ is taken at fixed $\unit n$. Since $\dot{\vec{R}}=\vec{v}$, the time derivative along the subhalo trajectory is $d/dt=\vec v\cdot\nabla_{\vec R}$. The initial line-of-sight jerk is therefore, 
\begin{equation}\label{eqn:static_doppler_green_derivative}
    j_\parallel(\unit{n}) \equiv  \dot{a}_\parallel(\unit{n}, 0)
    =
    GM(\vec{v}\cdot\nabla_{\vec{R}})\,
    \partial_L
    \left[
        \frac{1}{|\vec{R}-L\unit{n}|}
    \right].
\end{equation}
Writing the signal in this form allows us to use the standard multipole expansion:
\begin{equation}\label{eqn:static_doppler_green_multipole_expansion}
    \frac{1}{|\vec R-L\unit n|}
    =
    \sum_{\ell=0}^{\infty}
    \sum_{m=-\ell}^{\ell}
    \frac{4\pi}{2\ell+1}
    \frac{r_<^{\ell}}{r_>^{\ell+1}}\,
    Y_{\ell m}(\unit n)
    Y_{\ell m}^*(\unit R),
\end{equation}
where $r_<\equiv\min(R,L)$ and $r_>\equiv\max(R,L)$. The angular modes of the signal are then 
\begin{equation}\label{eqn:static_doppler_angular_mode_definition}
    j_{\parallel, \ell m}
    \equiv
    \int d^2\unit{n}\,
    Y_{\ell m}^*(\unit{n})\,
    j_{\parallel}(\unit{n}).
\end{equation}
 Using Eq.~\eqref{eqn:static_doppler_green_multipole_expansion} and Eq.~\eqref{eqn:static_doppler_green_derivative} this evaluates to
\begin{equation}\label{eqn:static_doppler_angular_mode_explicit}
    j_{\parallel, \ell m}
    =\frac{4\pi GM}{2\ell+1} 
    \left(\vec{v}\cdot\nabla_{\vec{R}}\right)
    \left[
    \partial_L\!\left(\frac{r_<^\ell}{r_>^{\ell+1}}\right)
    Y_{\ell m}^*(\unit{R})
    \right].
\end{equation}
To simplify our notation, we temporarily define
\begin{equation} \label{eqn:F_ell_definition}
    F_\ell\equiv\partial_L\!\left(\frac{r_<^\ell}{r_>^{\ell+1}}\right) = \begin{cases}
        \displaystyle -\frac{(\ell+1)R^\ell}{L^{\ell+2}},
        & R<L, \\[8pt]
        \displaystyle \frac{\ell L^{\ell-1}}{R^{\ell+1}},
        & R>L.
    \end{cases}
\end{equation}
For isotropic velocity directions, $\langle v_i v_j\rangle_{\unit{v}}=v^2\delta_{ij}/3$, so averaging $\left|j_{\parallel,\ell m}\right|^2$ over $\unit{v}$ gives
\begin{equation}\label{eqn:static_doppler_velocity_average}
    \left\langle
        \left|j_{\parallel,\ell m}\right|^2
    \right\rangle_{\unit{v}}
    =
    \frac{v^2}{3}
    \left(\frac{4\pi GM}{2\ell+1}\right)^2 \nabla_{\vec{R}}
        \left[F_\ell\, 
            Y_{\ell m}^*(\unit{R})
        \right] \cdot \nabla_{\vec{R}} 
        \left[F_\ell\, 
            Y_{\ell m}(\unit{R})
        \right] .
\end{equation}
Writing the gradient in spherical coordinates, $\nabla_{\vec R}=\unit R\,\partial_R+R^{-1}\nabla_{\unit{\Omega}}$, we evaluate
\begin{equation}
    \nabla_{\vec{R}} \left[F_\ell\,  Y_{\ell m}(\unit{R})\right] = \unit{R} \, (\partial_R F_\ell)\,  Y_{\ell m}(\unit{R}) + \frac{F_\ell}{R}\nabla_{\unit{\Omega}}Y_{\ell m}(\unit{R}) .
\end{equation}
The cross terms vanish, leaving
\begin{equation}%
    \left\langle
        \left|j_{\parallel,\ell m}\right|^2
    \right\rangle_{\unit{v}}
    =
    \frac{v^2}{3}
    \left(\frac{4\pi GM}{2\ell+1}\right)^2 \left[(\partial_R F_\ell)^2\,  |Y_{\ell m}(\unit{R})|^2 + \left(\frac{F_\ell}{R}\right)^2 |\nabla_{\unit{\Omega}}Y_{\ell m}(\unit{R})|^2
        \right] .
\end{equation}
We now average over the $2\ell+1$ values of $m$ to define $c_\ell^{\rm DM}$,
\begin{equation}\label{eqn:c_ell_DM_definition}
    c_\ell^{\rm DM}
    \equiv
    \frac{1}{2\ell+1}
    \sum_{m=-\ell}^{\ell}
    \left\langle
    \left|
         j_{\parallel, \ell m}
    \right|^2 \right\rangle_{\unit{v}} .
\end{equation}
This is evaluated using the spherical addition theorem and its angular derivative, 
\begin{equation}\label{eqn:static_doppler_harmonic_sums}
    \sum_m|Y_{\ell m}(\unit{R})|^2=\frac{2\ell+1}{4\pi}\implies 
    \sum_m|\nabla_{\unit{\Omega}} Y_{\ell m}(\unit{R})|^2
    =\ell(\ell+1)\frac{2\ell+1}{4\pi},
\end{equation}
to give   
\begin{equation}\label{eqn:c_ell_DM_2}
    c_\ell^\text{DM}
    =
    \frac{4\pi v^2}{3}
    \left(\frac{GM}{2\ell+1}\right)^2 \left[(\partial_R F_\ell)^2  + \frac{\ell(\ell+1)}{R^2} F_\ell^2
        \right].
\end{equation}
Note that the derivative $\partial_R F_\ell$ evaluates to
\begin{equation} \label{eqn:derivative_Fell}
    \partial_R F_\ell
    = 
    \begin{cases}
        \displaystyle \frac{\ell}{R}F_\ell,
        & R<L, \\[8pt]
        \displaystyle -\frac{\ell+1}{R}F_\ell,
        & R>L.
    \end{cases}
\end{equation}
Substituting Eq.~\eqref{eqn:F_ell_definition} and Eq.~\eqref{eqn:derivative_Fell} into Eq.~\eqref{eqn:c_ell_DM_2}, we have
\begin{align}\label{eqn:c_ell_DM_3}
    c_\ell^{\rm DM}
    =
   \frac{4\pi}{3}\left(\frac{GMv}{L^3}\right)^2\frac{\ell(\ell+1)}{2\ell+1} \times\begin{dcases*}
        (\ell+1)\left(\frac{R}{L}\right)^{2\ell-2} & if $R<L$, \\
        \ell\left(\frac{L}{R}\right)^{2\ell+4} & if $R>L$.
    \end{dcases*}
\end{align}
For a typical loudest flyby the subhalo lies close to the pulsar sphere, $|R-L|=r_\text{DM}\ll L$, so $R=L\mp r_\text{DM}$ in the two cases of Eq.~\eqref{eqn:c_ell_DM_3}. The powers of $R/L$ can then be approximated as exponentials
\begin{equation}
    \left(\frac{R}{L}\right)^{2\ell-2}=\left(1-\frac{r_\text{DM}}{L}\right)^{2\ell-2},\qquad
    \left(\frac{L}{R}\right)^{2\ell+4}=\left(1+\frac{r_\text{DM}}{L}\right)^{-(2\ell+4)},
\end{equation}
both of which reduce to $e^{-2\ell r_\text{DM}/L}$ for $\ell\gg1$. The two cases in Eq.~\eqref{eqn:c_ell_DM_3} then become
\begin{equation}\label{eqn:static_doppler_angular_spectrum_approx}
    \boxed{c_\ell^{\rm DM}\approx\frac{2\pi}{3}\left(\frac{GMv}{L^3}\right)^2\ell^2\,e^{-2\ell r_\text{DM}/L}, \quad \ell \gg 1.}
\end{equation}
For fixed $r_\text{DM}$, the spectrum grows as $c_\ell^{\rm DM}\propto\ell^2$ for $\ell\ll L/r_\text{DM}$, peaks near $\ell\simeq L/r_\text{DM}$, and is exponentially suppressed for $\ell\gg L/r_\text{DM}$. The weighted spectrum $\ell(\ell+1)c_\ell^{\rm DM}$ instead peaks near $\ell\simeq2L/r_\text{DM}$.

\section{GWB suppression in a joint multipulsar analysis}\label{supp:multi_pulsar}

In the main text we consider $\sim 100$ close pulsar pairs with separation of $d\sim 0.01-1$ pc. In Sec.~\ref{supp:potential_pairs}, we motivate this by taking an example of a globular cluster, namely Terzan 5, and showing that its expected MSP population alone could supply ${\sim}70$ independent pairs separated by less than $0.1\,\mathrm{pc}$, approaching this benchmark. The pulsars in a cluster like Terzan 5, however, are better modeled as a group of many close pulsars than as isolated pairs. We show numerically that the timing data from all these pulsars can be combined into channels with much stronger GWB suppression than the difference channel considered in the main text. In Sec.~\ref{supp:GWB_joint_suppress}, we explain this result analytically by showing that the GWB in many channels is suppressed to higher order than in the pairwise difference channel. However, as noted in the main text, the baryonic background in such a cluster would likely pose a significant challenge to using this suppression to detect a DM signal. Here, we only demonstrate GWB suppression and leave the calculation of the resulting DM sensitivity to future work.

\subsection{Numerical example: Terzan 5}
\label{supp:potential_pairs}

\begin{figure*}
    \centering
    \includegraphics[width=\linewidth]{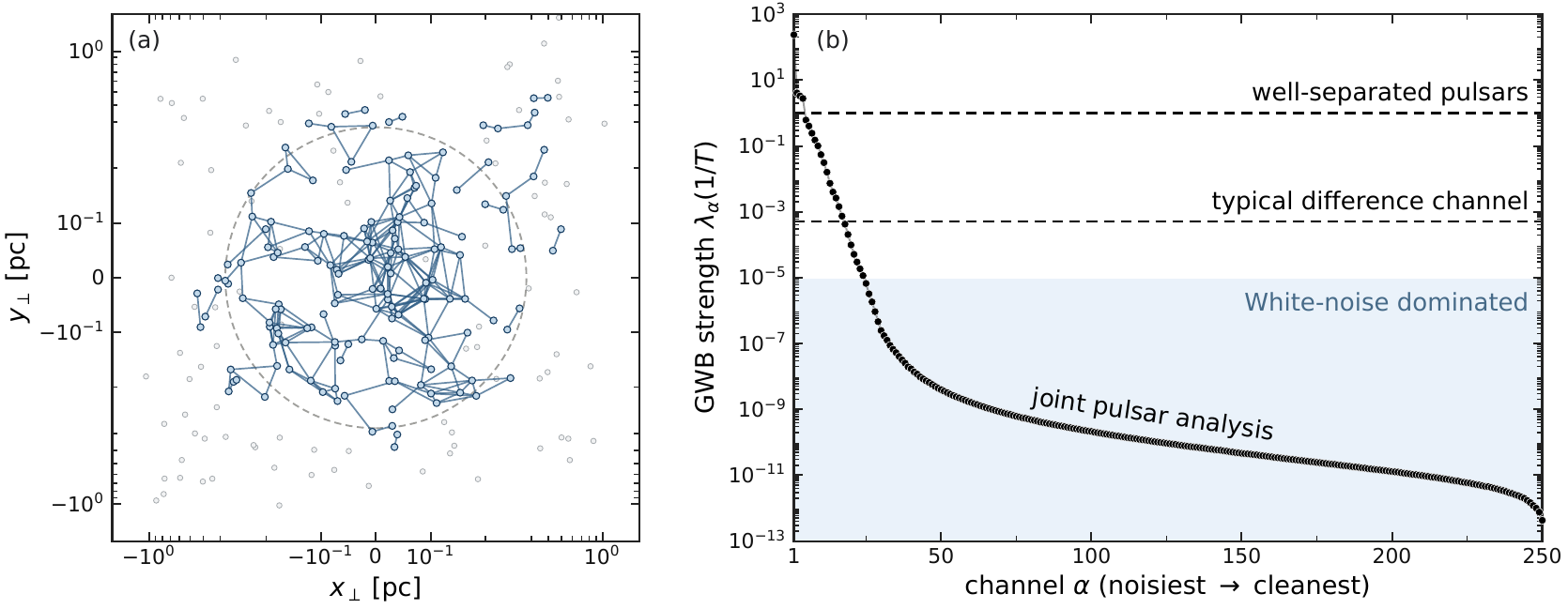}
    \caption{(a) A mock realization of Terzan 5 with $N_P=250$ pulsars (circles) drawn from Eq.~\eqref{eqn:terzan5_distribution}, shown as a sky projection. Edges link pulsars with three-dimensional separation below $0.1\,\mathrm{pc}$, and the dashed gray circle marks the core radius $r_c$. Many pulsars have several neighbors within $0.1\,\mathrm{pc}$, so the cluster is better modelled as a group of nearby pulsars rather than as isolated pairs. (b) The eigenvalues $\lambda_\alpha(f=1/T)$ of the $N_P\times N_P$ ORF matrix that quantify the GWB strength in each channel. The dots show their median over realizations of the Terzan~5 distribution in Eq.~\eqref{eqn:terzan5_distribution}. Although a typical difference channel in Terzan~5 (lower dashed line, Eq.~\eqref{eqn:lambda_typ}) substantially suppresses the GWB relative to the well-separated case (upper dashed line, $\lambda=1$), the residual GWB remains above the white-noise level. By contrast, a joint analysis of all $N_P=250$ pulsars concentrates most of the GWB power in a few channels and suppresses it much further in the rest, leaving most channels white-noise dominated (shaded blue). %
    }
    \label{fig:terzan5_joint_eigenspectrum}
\end{figure*}

We observe only the two-dimensional projection of a globular cluster. Let $r_\perp$ be the projected distance from the cluster center and $\Sigma(r_{\perp})$ the number of MSPs per unit area on the sky with that distance. Reference~\cite{urquhart2026newdeepradiocontinuum} considers $48$ precisely localized MSPs in Terzan 5, and fits their sky positions to a generalized King profile%
\begin{equation}\label{eqn:terzan5_surface_density}
    \Sigma(r_\perp)
    \propto
    \left[
        1+\left(\frac{r_\perp}{r_c}\right)^2
    \right]^{(1-3q)/2},
\end{equation}
finding $q=1.74\pm0.16$ and a core radius $r_c=9.6''\simeq0.275\,{\rm pc}$. %
Assuming spherical symmetry, we can reconstruct the three-dimensional distribution from the two-dimensional projection using the Abel inversion (see, \textit{e.g.}, Ref.~\cite{2010MNRAS.401.2433M}) %
\begin{equation}\label{eqn:Abel}
    p(r) \propto -%
    \int_r^{\infty}dr_{\perp}\,\frac{d\Sigma}{dr_{\perp}}\frac{1}{(r^2_{\perp}-r^2)^{1/2}} \, ,
\end{equation}
where $r$ is the physical distance from the cluster center and $p(r)$ is the probability density per unit volume for a single pulsar to be located at that distance. Evaluating the integral in Eq.~\eqref{eqn:Abel} using Eq.~\eqref{eqn:terzan5_surface_density}, we find
\begin{equation}\label{eqn:terzan5_distribution}
    p(r)
    \propto
    \left[
        1+\left(\frac{r}{r_c}\right)^2
    \right]^{-3q/2} \, .
\end{equation}
The same study also infers a population of at least $250$ radio-visible pulsars, of which only 49 have been discovered. Assuming that the undiscovered pulsars follow the same spatial distribution,\footnote{This estimate does not account for selection effects that may cause the brighter detected MSPs to be preferentially concentrated toward or away from the cluster core relative to the full population. %
} we draw $N_P=250$ positions from Eq.~\eqref{eqn:terzan5_distribution} and count pairs separated by less than $0.1\,\mathrm{pc}$. Across realizations, the $10$th--$90$th percentile ranges are $253$--$373$ total pairs and $68$--$78$ independent pairs, in which each pulsar belongs to at most one pair. Thus, Terzan~5 alone could already contain ${\sim}70$ independent pairs, approaching our $100$-pair benchmark. 

Notice, however, that there are many more total close pairs than independent close pairs, which indicates that many pulsars lie within $0.1$~pc of several other pulsars. As seen in Fig.~\ref{fig:terzan5_joint_eigenspectrum}(a), the cluster core is therefore better described as a group of $N$ pulsars inside a small sphere with radius $R\ll d_\text{coh}(f)$ rather than as a set of isolated pairs. 
Our analysis in the main text drops correlations of the GWB between distinct pairs. We now show numerically that 
retaining these correlations allows much stronger suppression of the GWB in a Terzan-5-like geometry.

The SNR from a joint analysis of $N_P$ pulsars is a generalization of Eq.~\eqref{eqn:pair_snr}:
\begin{equation}\label{eqn:general_N_snr}
    \mathrm{SNR}^2
    =
    \frac{4}{S_{\rm white}}
    \int_{1/T}^{\infty} df
    \sum_{\alpha=1}^{N_P}
    \frac{\left|\widetilde{\delta t}^{\,\perp}_{{\rm sig},\alpha}(f)\right|^2}{s_\alpha(f)} \, ,
\end{equation}
where $\lambda_{\alpha}(f)$ are the eigenvalues of the $N_P\times N_P$ ORF matrix $[\widehat{\mathbf{\Gamma}}(f)]_{ab} \equiv \Gamma^{\rm GWB}_{ab}(f)$, which characterize the GWB's contribution to the noise. The quantity $s_{\alpha}(f)\equiv 1+(f_\star/f)^{\gamma}\lambda_{\alpha}(f)$ is the total noise PSD of channel $\alpha$ normalized to $S_{\mathrm{white}}$, %
and $\alpha=1,2,\cdots,N_P$ labels the eigenvectors (or channels) of $\widehat{\mathbf{\Gamma}}(f)$. The main text considers a single pair ($N_P=2$), where the two eigenvalues, labeled by $\alpha =\pm$, are given by $\lambda_{\pm}=1\pm |\Gamma_{ab}|$. There we found that the GWB strength in the difference channel $\lambda_-$ is suppressed to $\mathcal{O}(d/d_\text{coh})^2$. For the Terzan~5 distribution of Eq.~\eqref{eqn:terzan5_distribution}, the median distance of a pulsar to its nearest neighbor is $d_\text{typ}\sim 0.08\, \mathrm{pc}$. For a typical Terzan~5 pair with separation $d=d_\text{typ}$, Eq.~\eqref{eqn:close_pulsar_ORF_schematic} gives the GWB strength in the difference channel as
\begin{equation} \label{eqn:lambda_typ}
    \lambda^\mathrm{typ}_-(f=1/T) = \frac{3}{40}\left(\frac{d_{\rm typ}}{d_{\rm coh}(1/T)}\right)^2.
\end{equation}
In Fig.~\ref{fig:terzan5_joint_eigenspectrum}(b), we compare this value with the $N_P=250$ eigenvalues $\lambda_\alpha(f=1/T)$ of the $N_P\times N_P$ ORF matrix, evaluated using the integral form in Eq.~\eqref{eqn:GWB_ORF_short_wavelength}. For the joint analysis, we show the median eigenvalue in each channel over realizations of the Terzan~5 distribution. The typical difference channel substantially suppresses the GWB but remains above the white-noise level. A joint analysis using the timing data of all $N_P$ pulsars in the cluster, on the other hand, concentrates most of the GWB power in a few channels and suppresses it much further in the rest, leaving most channels white-noise dominated. This occurs even though the typical nearest-neighbor separation is much larger than $d_\text{white}$. As we show in the next subsection, this is because many eigenvalues $\lambda_\alpha(f=1/T)$ in a joint analysis are suppressed to much higher order than $\mathcal{O}(R/d_\text{coh}(f))^2$.

\subsection{Higher-order suppression of the GWB}
\label{supp:GWB_joint_suppress}

To see how the joint analysis in Eq.~\eqref{eqn:general_N_snr} of $N_P$ pulsars within a sphere $R\ll d_{\rm coh}(f)$ improves the reach, we show below that the ORF of the GWB, $\widehat{\mathbf{\Gamma}}(f)$, has a large number of eigenvalues that are highly suppressed by powers of $(R/d_\text{coh})$. For instance, $\widehat{\mathbf{\Gamma}}(f)$ has at least $N_P-1$ eigenvalues suppressed to $\mathcal{O}(R/d_\text{coh})^2$, $N_P-4$ suppressed to $\mathcal{O}(R/d_\text{coh})^4$, $N_P-10$ suppressed to $\mathcal{O}(R/d_\text{coh})^6$, and so on. %
To derive this counting, we again assume that all the pulsars are at the same distance from the Earth, \textit{i.e.} $L_a \equiv L$. We choose the origin to coincide with the center of the sphere, denote the pulsar positions by $\vec{d}_a$, and define $\vec{\epsilon}_a\equiv \vec{d}_a/d_\text{coh}(f)$, with $|\vec{\epsilon}_a|\ll 1$. To leading order in $R/d_\text{coh}$, as in Eq.~\eqref{eqn:GWB_ORF_short_wavelength},
the ORF is given by  %
\begin{equation} \label{eqn:orf_epsilon}
    \Gamma^\text{GWB}_{ab}(f) \approx \frac{1}{2} + \frac{3}{2}\int \frac{{d^2 \unit{\Omega}}}{4\pi}  \sum_{A} \left[F^A_0(\unit{\Omega})\right]^2 e^{i\unit{\Omega}\cdot (\vec{\epsilon}_a-\vec{\epsilon}_b)} \, ,
\end{equation}
where $F^A_0(\unit{\Omega})$ is the antenna pattern given in Eq.~\eqref{eqn:antenna_pattern} evaluated for a pulsar along $\unit{n}_0$, the direction from Earth to the center of the sphere. 

We now count how many $\lambda_{\alpha}$ are suppressed to $\mathcal{O}(\epsilon^{2k})$ for some $k\geq 1$, where $\epsilon\equiv\max_a \lVert\vec{\epsilon}_a\rVert\ll 1$ is set by the farthest pulsar. This can be computed using the min-max theorem, which states that the number of eigenvalues of a Hermitian matrix that are at most $\delta$ equals the largest dimension of a subspace where $\vec{v}^{\dagger}\widehat{\mathbf{\Gamma}}\vec{v}\leq\delta\lVert\vec{v}\rVert^2$ for every vector $\vec{v}$ in it~\cite{Horn_Johnson_2012}. The number of $\lambda_{\alpha}$ that are $\mathcal{O}(\epsilon^{2k})$ is therefore at least the dimension of the subspace $\mathcal{V}_k$ of array-space vectors $\vec{v}$ for which $\vec{v}^{\dagger}\widehat{\mathbf{\Gamma}}\vec{v}$ begins at order $\epsilon^{2k}$, which we can determine without explicitly diagonalizing $\widehat{\mathbf{\Gamma}}$. We expand Eq.~\eqref{eqn:orf_epsilon} to write
\begin{equation}\label{eqn:vGammav}
    \vec{v}^{\dagger}\widehat{\mathbf{\Gamma}} \vec{v} = \frac{1}{2}\left| \sum_{a} v_{a} \right| ^2 + \frac{3}{2}\int \frac{{d^2 \unit{\Omega}}}{4\pi} \sum_{A}\left[ F_{0}^A(\unit{\Omega}) \right]^2\, \left| \sum_{a}e^{i\unit{\Omega}\cdot\vec{\epsilon}_{a}} \, v_{a} \right|^2.
\end{equation}
The dimension of $\mathcal{V}_k$ is determined by the number of constraints that the vectors $\vec{v}$ have to satisfy. Requiring $\vec{v}^{\dagger}\widehat{\mathbf{\Gamma}} \vec{v}$ to be bounded by $\mathcal{O}(\epsilon^{2k})$ implies that both terms in Eq.~\eqref{eqn:vGammav} are $\lesssim \mathcal{O}(\epsilon^{2k})$. The former requires $\sum_av_a=0$, which is a single constraint on $\vec{v}$. The latter requires $\sum_ae^{i\unit{\Omega}\cdot \vec{\epsilon}_a}v_a$ to be $\mathcal{O}(\epsilon^k)$ for every direction $\unit{\Omega}$. Expanding the phase for small $\epsilon$ gives 
\begin{equation}
    \sum_{a} e^{i\unit{\Omega}\cdot\vec{\epsilon}_{a}}\, v_{a}
    =
    \sum_{a} v_{a}
    + i\,\unit{\Omega}\cdot\sum_{a}\vec{\epsilon}_{a}\, v_{a}
    - \frac{1}{2}\sum_{a}\left(\unit{\Omega}\cdot\vec{\epsilon}_{a}\right)^2 v_{a}
    + \cdots \, ,
\end{equation}
where the $n$th term of the sum is $\mathcal{O}(\epsilon^n)$. We note here that the constraint from the $n=0$ term of this expansion %
is equivalent to requiring the first term in Eq.~\eqref{eqn:vGammav} to vanish. The sum can only be bounded by $\mathcal{O}(\epsilon^k)$ if every term with $n\leq k-1$ vanishes, and since $\unit{\Omega}$ is arbitrary, the corresponding moment must also vanish component by component 
\begin{equation}\label{eqn:phase_expand}
    \sum_{a}\epsilon_{a}^{i_{1}}\epsilon_{a}^{i_{2}} \cdots
    \epsilon_{a}^{i_{n}} \, v_{a} = 0,
    \qquad
    \forall\, i_{1},\dots,i_{n}\in \{\unit{x}, \unit{y}, \unit{z}\},
    \quad
    0 \leq n \leq k-1.
\end{equation}
At each order $n$, Eq.~\eqref{eqn:phase_expand} gives ${{n+2}\choose{2}}$ constraints. %
The dimension of $\mathcal{V}_k$ is then the dimension $N_P$ of the array space minus the number of linearly independent constraints. As the constraints above may not be independent, this gives us a lower bound on its dimension, %
\begin{equation}
    \text{dim}\, \mathcal{V}_k
    \geq N_P - \sum_{n=0}^{k-1} {{n+2}\choose{2}}
    =N_P - {{k+2}\choose{3}}.
\end{equation}
Evaluating this for $k=1,2,3,\dots$ gives the counting %
quoted above.

\section{Effect of degrading the timing benchmark} %
\label{supp:degrading_timing}

\begin{figure*}
    \centering
    \includegraphics[width=\linewidth]{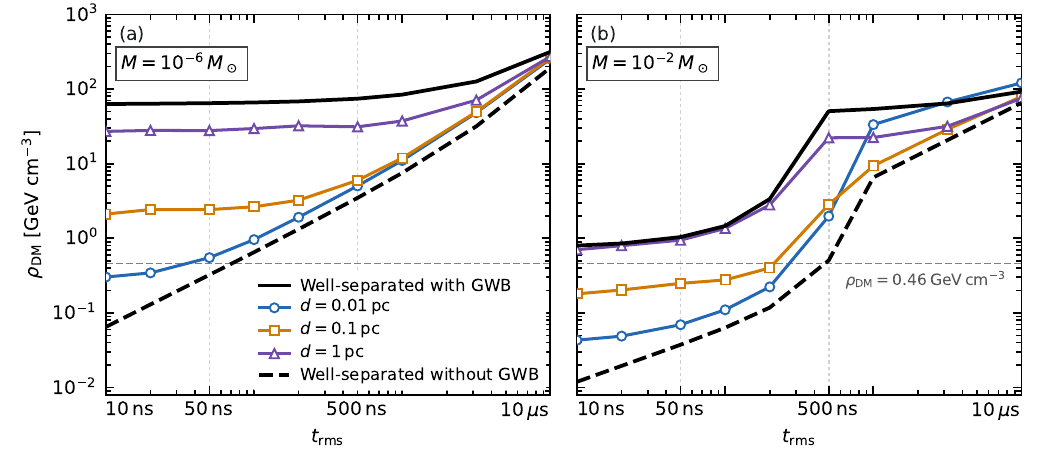}
    \caption{%
    The reach $\rho_\text{DM}$ for a close pulsar-pair analysis as a function of the white noise amplitude $t_\text{rms}$ from $10\, \rm{ns}$ to $10\, \rm{\mu s}$ at pair separations $d=0.01$, $0.1$, and $1\,\mathrm{pc}$. (a) $M = 10^{-6}M_\odot$, where the loudest event is dominated by the Doppler term. (b) $M = 10^{-2}M_\odot$, where the loudest event is instead dominated by the Shapiro delay. %
    The corresponding reach for a conventional well-separated array, with and without the GWB, is shown in black for reference.}
    \label{fig:timing_noise_degradation}
\end{figure*}

In this section, we discuss how much $t_\text{rms}$ (which we assume sets the white noise amplitude) can degrade before the close-pair advantage is lost. This is motivated by the fact that MSPs in environments that can host close pulsar pairs, such as globular clusters, are likely faint relative to their background, and the SKA-like timing precision $t_\text{rms}= 50\, \mathrm{ns}$ assumed in the main text may be challenging to achieve. 

For a conventional PTA analysis with well-separated MSPs, the reach of the DM signal does not worsen appreciably even when the timing precision is substantially degraded, %
for example up to $t_\text{rms}\sim1\,\mathrm{\mu s}$ at $M=10^{-6}\,M_{\odot}$. Whether close pairs retain this tolerance is not obvious, since their gain comes from removing the GWB, the very noise that made the conventional reach insensitive to white noise. In Fig.~\ref{fig:timing_noise_degradation}, we investigate how much the reach degrades for our close pair setup, at masses near our best sensitivity, $M= 10^{-6} M_\odot$ and $M=10^{-2}M_\odot$, where the DM signal is dominated by the Doppler and Shapiro term, respectively. At $M= 10^{-6}M_\odot$, the DM event is nearly in the dynamic limit, and its reach depends on the white noise moderately, weakening by a factor of $\sim 2-3$ when $t_\text{rms}$ is worsened from $50\, \mathrm{ns}$ to $500\, \mathrm{ns}$ for $d=0.1\, \mathrm{pc}$. At $M= 10^{-2} M_\odot$, the DM event is deep in the dynamic limit. The dependence of its reach on $t_\text{rms}$ is more complicated and stronger in this regime~\cite{Cherukupalli:2026cda}, and the same level of degradation in $t_{\mathrm{rms}}$ leads to an order of magnitude weakening in the reach. In both cases, close pairs retain their advantage over a conventional array with the same $t_\text{rms}$. Whether they also beat a conventional array timed to the benchmark $t_\text{rms}=50\,\mathrm{ns}$ depends on the signal: at $M=10^{-6}\,M_\odot$, where the Doppler term dominates, this holds to considerably larger $t_\text{rms}$, whereas at $M=10^{-2}\,M_\odot$, where the Shapiro delay dominates, it holds only for $t_\text{rms}\lesssim300\,\mathrm{ns}$.

In Ref.~\cite{Cherukupalli:2026cda}, we concluded that for a conventional PTA analysis using well-separated pulsars, reducing $t_\text{rms}$ brings diminishing returns as the GWB, unaffected by better timing precision, is the dominant noise source. We find that the same conclusion holds for the %
difference channel analysis on close pulsar pairs. Although the GWB's contribution is suppressed, it still eventually becomes the limiting noise as $t_\text{rms}$ improves, and the gain in sensitivity to DM saturates. %
Closer pairs have a weaker GWB in the difference channel, and thus this saturation starts setting in at a smaller $t_\text{rms}$, as shown in Fig.~\ref{fig:timing_noise_degradation}.

\putbib[biblio]
\end{bibunit}
\end{document}